\documentclass[twocolumn,prl,aps,notitlepage,showpacs,amsmath,amssymb,superscriptaddress]{revtex4-1}
\usepackage[latin9]{luainputenc}
\usepackage{amsmath}
\usepackage{amssymb}
\usepackage{color}
\usepackage{graphicx}
\usepackage{esint}



\begin{document}
\title{Semiclassical transport with Berry curvature: Chambers formula and applications to systems with Fermi surface topological transitions.}
\author{Emmanouil K. Kokkinis}
\affiliation{Department of Physics, Loughborough University, Loughborough, LE11
3TU, UK}
\affiliation{Physics department, University of Crete, Heraklion 71003, Greece.}
\author{Garry Goldstein}
\affiliation{Physics department, Boston University, Boston, Massachusetts 02215,
USA}
\author{Dmitri V. Efremov}
\affiliation{Department of Physics, Loughborough University, Loughborough, LE11
3TU, UK}
\affiliation{IFW Leibniz Institute for Solid State and Materials Research, Dresden, Helmholtz str. 20, 01069 Dresden, Germany.}
\author{Joseph J. Betouras}
\affiliation{Department of Physics, Loughborough University, Loughborough, LE11
3TU, UK}
\date{\today}
\begin{abstract}
Starting with general semiclassical equations of motion for electrons
in the presence of electric and magnetic fields, we extend the Chambers
formula to include in addition to a magnetic field, time-dependent
electric fields and bands with Berry curvature. We thereby compute
the conductivity tensor $\sigma_{\alpha\beta}\left(B,\omega\right)$
in the presence of magnetic field for bands in
two (2D) and three (3D) dimensions with Berry curvature. 
We focus then on
several applications to magnetotransport for metals with Fermi surface topological transitions in 2D. In particular, we consider a rectangular lattice and a model related to overdoped graphene, to investigate the signatures of different types of
Fermi surface topological transitions in metals in the Hall coefficient, Hall conductivity
$\sigma_{xy}$ and longitudinal conductivity $\sigma_{xx}$. The behavior
of those quantities as a function of frequency, when the electric
field is time dependent, is also investigated. As an example of non-zero Berry curvature, we study the magnetotransport of the Haldane model within this context. In addition, we provide
the linear and nonlinear electric current formula to order $E^{2}$. 
\end{abstract}
\maketitle

\section{\label{sec:Introduction}Introduction}

The majority of metals are well described by the Fermi liquid theory. Within this
formalism the classical Hall effect arises due to the electron path curvature in the presence of an external magnetic
fields \cite{Ziman_1972}. It is characterized by the Hall coefficient $R_{H} = E_y/(j_x B_z)$, where $j_x$ is the current density perpendicular to the applied magnetic field $B_z$ and $E_y$ is the induced electric field. In the case of a single band the Hall coefficient depends only on the sign of the charge and the density of the carriers. 
Therefore the Hall coefficient at zero temperature 
is widely used as a measure of the number of electrons or holes enclosed in a closed
orbit \citep{Ziman_1972, Abrikosov_1988}.
However, the Hall coefficient may deviate from the simple form of counting of the number of electrons or holes enclosed in the electron-and hole-like pockets. It happens, e.g.  in multi-band systems and in the systems close to the Fermi-surface topological transitions (FSTTs), as they are connected to singularities
in the density of states, called Van Hove singularities \citep{vanHove,Lifshitz_1960, Chandrasekaran, Chandrasekaran-Betouras}. 

There have been extensive studies of
Lifshitz transitions (pocket appearing/disappearing or neck formation/collapse)
and associated Van Hove singularities in a variety of materials including
cuprates, iron based superconductors, cobaltates, Sr$_{2}$RuO$_{4}$ and
heavy fermions  \citep{Aoki,Barber,Bernhabib,Khan,Coldea,Okamoto,Sherkunov-Chubukov-Betouras,Slizovskiy-Chubukov-Betouras,Stewart,Yelland}.
Most of these materials show logarithic-type Van Hove singularity, which corresponds to a logarithmic singularity of the density of states at the Lifshitz transition in 2D. However, there is a recent surge of interest in
higher order Van Hove singularities, which manifest itself in the stronger than logarithmic singularity in the density of states. These are the result of more complicated FSTTs.
A prominent example is Sr$_{3}$Ru$_{2}$O$_{7}$,
where a $n=4$ Van Hove singularity connected to a more complicated FSST, is shown to exist in the presence
of an external magnetic field \citep{Efremov_2019}. Higher
order Van Hove saddle points has been observed  in highly overdoped graphene
and twisted bilayer graphene \cite{Rosenzweig_2020, Yuan_Isobe_Fu, Isobe_Fu, LiangFu} and may be highly relevant for the recently observed phases of Bernal bilayer graphene \cite{Zhou_2021, Shtyk_2017}. 

In this work, we study the behavior of the Hall coefficient across a FSTT that correspond to a high-order Van Hove singularities using the Boltzman equation in the presence of a static magnetic field, a Berry curvature ${\bf \Omega}\left({\bf k}\right)$ and a low-frequency electric
field ${\bf E}\left({\bf r},t\right)={\bf E} ({\bf r}) \exp\left(i\omega t\right)$.   
The solution of the Boltzmann equation is provided in form of the Chambers formula \cite{Chambers_1952}, which is widely used in studies of the magnetotransport   \cite{Callaway_1991,Shockley_1950,Kittel_1987,Singh_2017,Ziman_1960,Ziman_1972,Jones_1973,Wilson_1953,Galperin,Abrikosov_1988,Quinn_2009}.
We extend the Chambers formula to the case of Berry curvature.
%

The equations of motion used in this work
are correct to leading order in the electric field $E$ and magnetic
field $B$, but remarkably, retain the same form in corrected to order $E^{2}$ and $B^{2}$ \cite{Gao_2019, Gao_PRL14}.
However, using the methods presented here
it is possible to study the Boltzmann equation for arbitrary
magnetic fields and electric fields with sufficiently accurate equations
of motion \cite{Gao_2019}. To illustrate this idea we use the
leading order equations of motion to study the solution to the
Boltzmann equation in the magnetic field $B$ and to
quadratic order in the electric field $E^{2}$ (within the leading order equations of motion).
We ignore the Zeeman
splitting and consider spinless fermions thereby reducing to the case
of negligible spin orbit coupling \cite{Gao_2019,Gao_PRL14, Xia_2005,Xia_2006,Xiao_2009}, thus avoiding more complicated expressions \cite{Vanderbilt}.
We also neglect any changes to the chemical potential due to the magnetic
field, these can be incorporated straightforwardly \cite{Gao_2019, Gao_PRL14, Xia_2005,Xia_2006,Xiao_2009}.

The main contributions of this work are: 
(i) short and clear derivations of the Boltzmann equation relevant
to the semiclassical motion of Fermi liquids, including the Berry
curvature for leading order equations of motion. (ii) explicit solution
of that Boltzmann equation to all orders in the magnetic and electric
fields (formally exact), which can be implemented numerically for leading order equations
of motion. (iii) method for semi-analytical asymptotic expansions
of the Boltzmann equation solution to all orders in magnetic field
and to an arbitrary (finite) order in the electric field for any set
of equations of motion. (iv) explicit expansion of Boltzmann equation
solution to linear and quadratic orders in the electric field with Berry curvature
using the developed method, thus obtaining linear and bilinear response (recently introduced in Ref. \citep{Sodemann_2015,Facio_2018}),  using the
leading order equations of motion. 

It is worth emphasizing that the work is valid within the different Fermi surface topologies but away from the transition points.
At the transition points, as the Fermi velocity approaches zero, quantum effects become important as well as possible out-of-equilibrium effects in a time-dependent electric field. In particular magnetic breakdown quantum description of field-induced tunneling between semiclassical orbits shall be considered carefully close to Van Hove singularities \cite{Falicov, Glazman}. We, therefore, restrict the work away of those points.

The rest of the paper is organized as follows: in Section II 
we review the semiclassical equations of motion in the presence of
external fields and Berry curvature, as well as the Boltzmann equation
that follows. In Section III 
we derive the new 2D and 3D version of the Chamberss formula with
Berry curvature and time dependent external electrical field. In Sections
IV-VI 
we apply these results to specific classes of materials. 
t
In Section
\ref{sec:Application:-regular-van} we discuss
the case of a Lifshitz transition of the form neck formation/collapse verifying that the rapid
change between electron like to hole like Fermi surfaces leads to
a rapid change of the Hall coefficient, which we also study as a function
of frequency. In Section \ref{subsec:Application:-supermetal-saddle}
we present the application to FSTT that corresponds to a higher Van Hove saddle in graphene
and show how different FS topologies lead
to different responses in conductivities. In Section \ref{sec:Hall-conductivity-of}
we study the Hall conductivity of the Haldane model, as a representative example
of a system with Berry curvature.
In Section \ref{sec:Conclusion} we conclude. 

\section{\label{sec:Derivation-of-Eq.-1}Semiclassical equations of motion
- Boltzmann equation and solution}

The semiclassical equations of motion for electrons of a single particle
Hamiltonian $H\left(k\right)$ to leading order in electric and magnetic
fields was discussed in numerous works \citep{Deyo_2009,Sundaram_1999,Xia_2005,Xiao_2009,Gao_2019,Xia_2006,Chang_1995,Chang_1996,Chang_2008} and can be written as:
\begin{align}
\frac{d{\bf r}}{dt} & =D^{-1}\left({\bf r},{\bf k},t\right)\left[{\bf \nabla_{k}}\varepsilon_{M}\left({\bf k}\right)+e{\bf E}\left({\bf r},t\right)\times{\bf \Omega}\left({\bf k}\right)\right.\nonumber \\
 & \left.+e\left({\bf \Omega}\left({\bf k}\right)\cdot{\bf \nabla_{k}}\varepsilon_{M}\left({\bf k}\right)\right){\bf B}\left({\bf r},t\right)\right], \label{eq:Vector_ODE-1}\\
\frac{d{\bf k}}{dt} & =-D^{-1}\left({\bf r},{\bf k},t\right)\left[e{\bf E}\left({\bf r},t\right)+e{\bf \nabla_{k}}\varepsilon_{M}\left({\bf k}\right)\times{\bf B}\left({\bf r},t\right)\right.\nonumber \\
 & \left.+e^{2}\left({\bf B}\left({\bf r},t\right)\cdot{\bf E}\left({\bf r},t\right)\right){\bf \Omega}\left({\bf k}\right)\right]. \label{eq:Vector_ODE-2}
\end{align}
Here we have introduced $D^{-1}\left({\bf r},{\bf k},t\right)=\frac{1}{1+e{\bf B}\left({\bf r},t\right)\cdot{\bf \Omega}\left({\bf k}\right)}$.
The Berry curvature ${\bf \Omega}\left({\bf k}\right)$ is defined as the pseudo-vector ${\bf \Omega}\left({\bf k}\right)={\bf \nabla_{k}}\times{\cal A}({\bf k})$,
where ${\cal A}({\bf k})=i\left\langle u({\bf k})\right|{\bf \nabla_{k}}\left|u({\bf k})\right\rangle $
is the Berry connection and $\left|u({\bf k})\right\rangle $ an eigenstate
of the Hamiltonian $H\left({\bf k}\right)$.
The electron dispersion up to the second order of the magnetic fields can be written as \citep{Xia_2005,Xiao_2009}: 
$$\varepsilon_{M}\left({\bf k}\right)=\varepsilon\left({\bf k}\right)-{\bf m}\left({\bf k}\right)\cdot{\bf B}$$
with: 
\begin{equation}
{\bf m}\left({\bf k}\right)=-i\frac{e}{2\hbar}\left\langle {\bf \nabla_{k}}u\right|\times\left[H\left({\bf k}\right)-\varepsilon\left({\bf k}\right)\right]\left|{\bf \nabla_{k}}u\right\rangle . \label{eq:Magnetization-1-1}
\end{equation}
For simplicity, we consider below two dimensional (2D) case with both the electric and magnetic fields being uniform.  Furthermore we consider time-independent
magnetic field along z-axis  (which in 2D is out of plane). The electric field is taken weakly time-dependent, that we can neglect the induction of the magnetic field. 
In two dimensions (2D), the Berry curvature is perpendicular to the plane 
$\Omega_{z}\left({\bf k}\right)=\partial_{k_{x}}{\cal A}_{y}-\partial_{k_{y}}{\cal A}_{x}=-2Im\left\langle \partial_{k_{x}}u\mid\partial_{k_{y}}u\right\rangle $ and   ${\bf \Omega}\left({\bf k}\right)\cdot{\bf \nabla_{k}}\varepsilon_{M}\left({\bf k}\right)=0$. This configuration considerably
simplifies the set of Eqs. (\ref{eq:Vector_ODE-1}-\ref{eq:Vector_ODE-2}). 

Now  we  derive of the Boltzmann equation for our case. The derivation is the same in 2D or 3D so
the dimensionality will be denoted by $d$ and we will distinguish
the differences in the next subsections. Let us take a small volume
in phase space then the total number of particles in the system satisfies
the continuity equation \citep{Kim_2014,Gao_2019,Arovas_2021,Deyo_2009}:
\begin{equation}
\frac{\partial[Df]}{\partial t}+\nabla\cdot\left(\mathbf{w}Df\right)=\frac{D}{\tau_{S}({\bf k})}\left(f_{0}-f\right),
\end{equation}
where $f=f\left(\mathbf{k},\mathbf{r},t\right)$ is the distribution of electrons while $f_{0}=f_0\left(\varepsilon_M(\mathbf{k})\right)$
is the Fermi-Dirac distribution, $\tau_{S}({\bf k})$ is the relaxation
time  and $\mathbf{w}=\left(\frac{d{\bf r}}{dt},\frac{d{\bf k}}{dt}\right),\:\nabla=\left(\nabla_{\mathbf{r}},\nabla_{\mathbf{k}}\right)$
is the six velocity in phase space. Using the identity \citep{Arovas_2021,Xia_2005}
\begin{eqnarray}
\nabla\cdot\mathbf{w} & = & -\frac{d}{dt}\ln\left(D\left(\mathbf{k},\mathbf{r},t\right)\right)\nonumber \\
 & = & -\frac{\partial}{\partial t}\ln\left(D\left(\mathbf{k},\mathbf{r},t\right)\right)-\mathbf{w}\cdot\nabla\ln\left(D\left(\mathbf{k},\mathbf{r},t\right)\right)\label{eq:Log_derivative}
\end{eqnarray}
we obtain the Boltzmann equation for the distribution of electrons:
\begin{equation}
\frac{\partial}{\partial t}[f\left(\mathbf{k},\mathbf{r},t\right)]+\mathbf{w}\cdot\nabla\left[f\left(\mathbf{k},\mathbf{r},t\right)\right]=\frac{\left[f_{0}\left(\varepsilon_{M}\right)-f\left(\mathbf{k},\mathbf{r},t\right)\right]}{\tau_{S}({\bf k})}\label{eq:Boltzmann}
\end{equation}
with $\frac{d{\bf k}}{dt}$ and $\frac{d{\bf r}}{dt}$ being given
by Eq. (\ref{eq:Vector_ODE-2}). By considering spatially uniform
magnetic and electric fields, the equation become: 
\begin{equation}
\frac{\partial f}{\partial t}+\frac{d{\bf k}}{dt}\frac{df}{d{\bf k}}=-\frac{1}{\tau_{S}({\bf k})}\left(f-f_{0}\left({\bf \varepsilon_{M}}\right)\right)\label{eq:Boltzmann-1}.
\end{equation}
The solution to Eq. (\ref{eq:Boltzmann-1}) is given by \citep{MacCallum_1963}:
\begin{align}
 & f\left({\bf k},t\right)=f\left({\bf k_{0}},t_{0}\right)\exp\left(-\int_{t_{0}}^{t}\frac{ds}{\tau_{S}(\mathbf{k}\left(s\right))}\right)\nonumber \\
 & +\int_{t_{0}}^{t}ds \frac{f_{0}\left(\varepsilon_{M}\left({\bf k}(s)\right)\right)}{\tau_{S}\left({\bf k}(s\right))}\exp\left(-\int_{s}^{t}\frac{dt'}{\tau_{S}\left({\bf k}(t'\right))}\right)\label{eq:Solution-2}.
\end{align}
In the above $f\left({\bf k}_{0},t_{0}\right)$ corresponds to the initial
conditions. The components of the current density are given by: 
\begin{equation}
J_{\alpha}\left(t\right)=-e\int D\left({\bf k}\right)\frac{d^{d}k}{\left(2\pi\right)^{d}}\frac{dr_{\alpha}}{dt}f\left({\bf k},t\right),\label{eq:Solution-4}
\end{equation}

\noindent where the volume $V$ of the unit cell of the lattice is
set to unity and $d=2,3$ is the dimensionality of the system. If
the initial time is set to $t_0\rightarrow-\infty$ then: 
\begin{equation}
f\left({\bf k},t\right)\!=\!\!\int_{-\infty}^{t}\!\!\!\!\!\!ds\frac{f_{0}\left(\varepsilon_{M}\!\left({\bf k}(s)\right)\right)}{\tau_{S}\left(\mathbf{k}\left(s\right)\right)}\!\exp\!\left(\!-\!\int_{s}^{t}\!\!\!\frac{dt'}{\tau_{S}\left(\mathbf{k}\left(t'\right)\right)}\!\right)\!\label{eq:Density-2}
\end{equation}
and we note if we further linearize Eq. (\ref{eq:Density-2}) we recover
the linear response equations. 

\section{\label{subsec:Main-resuls}General results}

We first present here the general formulae, while the derivations
are left for the Appendix \ref{sec:Main-calculation}. The 2D and
3D cases are treated separately, due to the differences in the equations
of motion Eq. (\ref{eq:Vector_ODE-2}), when the Berry curvature is
taken into account. As stressed in the introduction, the formulae
are exact within the accuracy of the leading order equations of motion in the fields $E$ and $B$.

\subsection{\label{subsec:3D-case}Three-dimensional case}

To find $\mathbf{k}\left(t\right)$, the equation of motion is given
by Eq.(\ref{eq:Vector_ODE-2}). In the limit where $\mathbf{E}\left(t\right)$
is small, the last term of Eq.(\ref{eq:Vector_ODE-2}) can be taken
as a perturbation. In that spirit let us denote by $\mathbf{k}_{0}\left(t\right)$
the solution to the equation: 
\begin{equation}
\frac{d\mathbf{k}_{0}\left(t\right)}{dt}=-eD^{-1}\left(\mathbf{k}_0\right)\left[{\bf \nabla_{k}}\varepsilon_{M}\left({\bf k}\left(t\right)\right)\times{\bf B}\right]_{\mathbf{ k=k_0}}\label{eq:Zero_olution-1}
\end{equation}
and write: $\mathbf{E}\left(t\right)=\lambda\mathbf{E}\left(t\right)$
with $\lambda=1$. Then Eq.(\ref{eq:Vector_ODE-2}) becomes analytically
dependent on the parameter $\lambda$ and its solutions can be written
as an analytic asymptotic series of the form \citep{Young_2017,Bauer_2015,Nayfeh_1973}:
\begin{equation}
\mathbf{k}\left(t\right)=\mathbf{k}_{0}\left(t\right)+\lambda\mathbf{k}_{1}\left(t\right)+\lambda^{2}\mathbf{k}_{2}\left(t\right)+....\label{eq:Solution-3}
\end{equation}
with $\mathbf{k}_{1}\left(t_{0}\right)=\mathbf{k}_{2}\left(t_{0}\right)=\mathbf{k}_{3}\left(t_{0}\right)=...=0.$. These expansions and the method below are valid to any accuracy with respect to the external fields in the equations of motion.
To order $\lambda$: 
\begin{eqnarray}
\nonumber
 && \frac{d{{\bf k}}_{1}\left(t\right)}{dt}= \\
 \nonumber
 && -e \sum_\beta k_{1\beta}(t) \left[\frac{\partial}{\partial k_\beta}
 \left(D^{-1}\left(\mathbf{k}\right)\left[{\bf \nabla_{k}}\varepsilon_{M}\left({\bf k}\left(t\right)\right)\times{\bf B}\right] \right) \right]_{\mathbf{k}=\mathbf{k}_0} \\
 && -eD^{-1}\left(\mathbf{k}_{0}\left(t\right)\right)\left[{{\bf E}}\left(t\right)+e\left({\bf B}\cdot{\bf E}\left(t\right)\right){{\bf \Omega}}\left({\bf k}_{0}\left(t\right)\right)\right]\label{eq:Derivative-1}. 
\end{eqnarray}

Taking into account that $\frac{d\varepsilon\left(\mathbf{k}\left(t\right)\right)}{dt}=\nabla_{\mathbf{k}}\varepsilon\left(\mathbf{k}\left(t\right)\right)\cdot\frac{d\mathbf{k}}{dt}$,
we can also write perturbatively: 
\begin{equation}
\varepsilon_{M}\left(t\right)=\varepsilon_{0}\left(t\right)+\lambda\varepsilon_{1}\left(t\right)+\lambda^{2}\varepsilon_{2}\left(t\right)+...\label{eq:Transform}
\end{equation}
with $\varepsilon_0(t_0)=\varepsilon_M(t_0)$ and $\varepsilon_{1}\left(t_{0}\right)=\varepsilon_{2}\left(t_{0}\right)=\varepsilon_{3}\left(t_{0}\right)=...=0$. Again this is valid for any accuracy of the equations of motion.
Then to order $\lambda$ and for $t>t_{0}$:
\begin{eqnarray*}
 &&\varepsilon_{1}\left(t\right)=e\int_{t_{0}}^{t}dsD^{-1}\left(\mathbf{k}_{0}\left(s\right)\right)\nonumber \\
 && \times  \left[\nabla_{\mathbf{k}}\varepsilon_{M}\left(\mathbf{k}\left(s\right)\right)\cdot\left({\bf E}\left(s\right)+\left({\bf B}\cdot{\bf E}\left(s\right)\right){\bf \Omega}\left({\bf k}\left(s\right)\right)\right)\right]_{\mathbf{k=k_0}}
 \label{eq:Solution-1-2-1-1}
\end{eqnarray*}
and to order $\lambda^{2}$ 
\begin{eqnarray}
 &  & \varepsilon_{2}\left(t\right)=e\int_{t_{0}}^{t}ds\sum_{\alpha\beta}{E}_{\beta}\left(s\right){k}_{1\alpha}\left(s\right)\frac{\partial}{\partial{k}_{\alpha}}
 \Bigg[D^{-1}\left(\mathbf{k}_{0}\left(s\right)\right)  \nonumber \\
 \nonumber
 &  & \left. \times\left[\frac{\partial}{\partial{k}_{\beta}}\varepsilon_{M}\left(\mathbf{k}\left(s\right)\right)+e{B}_{\beta}\left(\nabla_{\mathbf{k}}\varepsilon_{M}\left(\mathbf{k}\left(s\right)\right)\cdot{\bf \Omega}\left({\bf k}\left(s\right)\right)\right)\right]\right]_{\mathbf{k=k_0}}\label{eq:E_2-1-1}
\end{eqnarray}

We introduce for convenience the quantities: 
\begin{eqnarray}
\nonumber
{\bf u_{0}}({\bf k})&\equiv&D^{-1}\left({\bf k}\right)\left[\nabla_{{\bf k}}\varepsilon_{M}\left({\bf k}\right)+e\left({\bf \Omega}\left({\bf k}\right)\cdot\nabla_{{\bf k}}\varepsilon_{M}\left({\bf k}\right)\right){\bf B}\right]
\label{eq:velocity}  \\
\nonumber
\eta(t;t_0)&\equiv&\exp\left(-\int_{t_{0}}^{t}\frac{ds}{\tau_{S}\left(\mathbf{k}_{0}\left(s\right)\right)}\right)
\end{eqnarray}
As a result, we obtain the current to linear order in $E$: 
\begin{widetext}
\begin{eqnarray}
\nonumber
J_{\alpha}^{\left(1\right)}\left(t\right)&=& - e^{2}\int\frac{d^{3}k}{\left(2\pi\right)^{3}}\left[\mathbf{E}\left(t\right)\times{\bf \Omega}\left(\mathbf{k}\right)\right]_{\alpha}f_{0}\left(\varepsilon_{M}\left(\mathbf{k}\left(t\right)\right)\right)-e\int\frac{d^{3}k}{\left(2\pi\right)^{3}}D\left(\mathbf{k}\right){u}_{{0}\alpha}\left(\mathbf{k}\right)\frac{\partial f_{0}\left(\varepsilon_{M}\left(\mathbf{k}_{0}(t)\right)\right)}{\partial\varepsilon}\times\\
 &\times & \int_{-\infty}^{t}D^{-1}\left(\mathbf{k}_{0}\left(t'\right)\right)\nabla_{\mathbf{k}}\varepsilon\left(\mathbf{k}_{0}\left(t'\right)\right)\cdot\left[\mathbf{E}\left(t'\right)+e\left({\bf B}\cdot{\bf E}\left(t'\right)\right){\bf \Omega}\left({\bf k}_{0}\left(t'\right)\right)\right] \eta(t;t') dt'\label{eq:Current-2-1-1}
\end{eqnarray}
\end{widetext}
where the first term generalizes the Streda formula \citep{Xiao_2009},
while the second term generalizes the Chambers formula \citep{Chambers_1952}.
Next, to order $E^{2}$ we obtain: 
\begin{widetext}
\begin{align}
 & J_{\alpha}^{\left(2\right)}\left(t\right)=-e\int\frac{d^{3}k}{\left(2\pi\right)^{3}}D\left(\mathbf{k}\right){u}_{{0}\alpha}\left(\mathbf{k}\right) \times \nonumber \\
 &  \left[e\frac{\partial f_{0}\left(\varepsilon\left(\mathbf{k}_{0}(t)\right)\right)}{\partial\varepsilon}\int_{-\infty}^{t}dt'\sum_{\gamma}{k}_{1\gamma}\left(t'\right)\frac{\partial}{\partial{k}_{\gamma}}\left[D^{-1}\left(\mathbf{k}_{0}\left(t'\right)\right)\left[\nabla_{\mathbf{k}}\varepsilon\left(\mathbf{k}_{0}\left(t'\right)\right)\cdot\left[\mathbf{E}\left(t'\right)+e\left({\bf B}\cdot{\bf E}\left(t'\right)\right){\bf \Omega}\left({\bf k}_{0}\left(t'\right)\right)\right]\right]\right]   \eta(t;t')   \right.\nonumber \\
 & +e^2\frac{\partial^{2}f_{0}\left(\varepsilon\left(\mathbf{k}_{0}(t)\right)\right)}{\partial\varepsilon^{2}}\int_{-\infty}^{t}dt'D^{-1}\left(\mathbf{k}_{0}\left(t'\right)\right)\nabla_{\mathbf{k}}\varepsilon\left(\mathbf{k}_{0}\left(t'\right)\right)\cdot\left[\mathbf{E}\left(t\right)+e\left({\bf B}\cdot{\bf E}\left(t'\right)\right){\bf \Omega}\left({\bf k}_{0}\left(t'\right)\right)\right]  \eta(t;t') \times\nonumber \\
 & \times\int_{t'}^{t}dlD^{-1}\left(\mathbf{k}_{0}\left(l\right)\right)\nabla_{\mathbf{k}}\varepsilon\left(\mathbf{k}_{0}\left(l\right)\right)\cdot\left[\mathbf{E}\left(l\right)+e\left({\bf B}\cdot{\bf E}\left(l\right)\right){\bf \Omega}\left({\bf k}_{0}\left(l\right)\right)\right]\nonumber \\
 & \left.+e\frac{\partial f_{0}\left(\varepsilon\left(\mathbf{k}_{0}(t)\right)\right)}{\partial\varepsilon}\int_{-\infty}^{t}dt'D^{-1}\left(\mathbf{k}_{0}\left(t'\right)\right)\nabla_{\mathbf{k}}\varepsilon\left(\mathbf{k}_{0}\left(t'\right)\right)\cdot\left[\mathbf{E}\left(t'\right)+e\left({\bf B}\cdot{\bf E}\left(t'\right)\right){\bf \Omega}\left({\bf k}_{0}\left(t'\right)\right)\right]  \eta(t;t')
 \int_{t'}^{t}dl\frac{\nabla_{\mathbf{k}}\tau_{S}\left(\mathbf{k}_{0}(l)\right)\cdot\mathbf{k}_{1}(l)}{\tau_{S}^{2}\left(\mathbf{k}_{0}\left(l\right)\right)}\right]\nonumber \\
 & -e^{3}\int\frac{d^{3}k}{\left(2\pi\right)^{3}}\left[\mathbf{E}\left(t\right)\times{\bf \Omega}\left(\mathbf{k}\right)\right]_{\alpha}\frac{\partial f_{0}\left(\varepsilon\left(\mathbf{k}\right)\right)}{\partial\varepsilon}\int_{-\infty}^{t}dt'D^{-1}\left(\mathbf{k}_{0}\left(t'\right)\right)\nabla_{\mathbf{k}}\varepsilon\left(\mathbf{k}_{0}\left(t'\right)\right)\cdot\left[\mathbf{E}\left(t'\right)+e\left({\bf B}\cdot{\bf E}\left(t'\right)\right){\bf \Omega}\left({\bf k}_{0}\left(t'\right)\right)\right]   \eta(t;t') \label{eq:Current_E^2-3-2}
\end{align}
\end{widetext}

The first term is a shift of the Fermi-Dirac distribution due to the
electric field, the second term is the quadratic shift, the third
term is a novel one that corresponds to a non-constant relaxation
time and the last term generalizes the Berry dipole introduced in
\citep{Sodemann_2015,Facio_2018} to potentially any magnetic field and electric
fields with arbitrary time dependence.  As it is explained in Appendix D,  it is important in order convergence to be guaranteed, that the eigenvalues of the matrix:
\begin{eqnarray}
\nonumber
&&M_{\alpha\beta}\left(s\right)=\\
\nonumber
&&-e\sum_{\gamma\delta}\varepsilon_{\alpha\gamma\delta}{B_{\delta}}\frac{\partial}{\partial{k}_{\beta}}\left[D^{-1}\left(\mathbf{k}_{0}\left(s\right)\right)\left[\frac{\partial}{\partial{k}_{\gamma}}\varepsilon_{M}\left({\bf k}_{0}\left(s\right)\right)\right]\right]
\end{eqnarray}
is smaller than $\frac{1}{\tau_{S}\left(\mathbf{k}\right)}$.

\subsection{\label{sec:Berry-Curvature} Two-dimensional case }

We can now reduce all previous results to 2D. The equation of motion
reads: 
\begin{equation}
\frac{d{\bf k}}{dt}=-D^{-1}\left({\bf r},{\bf k},t\right)\left(e{\bf E}\left(t\right)+e\nabla_{{\bf k}}\varepsilon_{M}\left({\bf k}\right)\times{\bf B}\right)\label{eq:ODE_general-1-1}
\end{equation}
If we introduce for convenience the notation: 
\begin{equation}
\mathbf{u_{0}^{2D}}\left(\mathbf{k}\right)=D^{-1}\left(\mathbf{k}\right)\nabla_{{\bf k}}\varepsilon_{M}\left({\bf k}\right),
\end{equation}
we arrive at a  similar expression for the current as Eq.(\ref{eq:Current-2-1-1})
and (\ref{eq:Current_E^2-3-2}) with the only differences that ${\bf u_{0}}$
is replaced by $\mathbf{u_{0}^{2D}}$, the integration over all ${\bf k}$'s
is now a 2D integration and ${\bf B}\cdot{\bf E}\left(t\right)\rightarrow0$.
The result for the linear response is:
\begin{align}
&J_{\alpha}^{\left(1\right)}\left(t\right) =-e^{2}\int\frac{d^{2}k}{\left(2\pi\right)^{2}}\left[\mathbf{E}\left(t\right)\times{\bf \Omega}\left(\mathbf{k}\right)\right]_{\alpha}f_{0}\left(\varepsilon_{M}\left(\mathbf{k}\left(t\right)\right)\right)\nonumber\\
&-e^{2}\int\frac{d^{2}k}{\left(2\pi\right)^{2}}D\left(\mathbf{k}\right)\mathbf{u}_{\mathbf{0}\alpha}^{\mathbf{2D}}\left(\mathbf{k}\right)\frac{\partial f_{0}\left(\varepsilon_{M}\left(\mathbf{k}_{0}(t)\right)\right)}{\partial\varepsilon} \times \nonumber\\
&\times \int_{-\infty}^{t}D^{-1}\left(\mathbf{k}_{0}\left(t'\right)\right)\nabla_{\mathbf{k}}\varepsilon\left(\mathbf{k}_{0}(t')\right)\cdot\mathbf{E}\left(t'\right)\eta(t;t')dt'\label{eq:Current-2-1-2}
\end{align}
For completeness we include the formula for the nonlinear response in 2D in the Appendix
(\ref{subsec:Main-calculation-1}) and we proceed to the scaling analysis
below.

\subsubsection{\label{sec:Scaling} Scaling analysis in 2D}

From the expression of the current we obtain the expression of the
conductivity which is divided into a part that we call topological
and a regular part.
\begin{widetext}
\begin{align}
 & \sigma_{\alpha\beta}\left(e,\tau_{S},T,\mathbf{B},\omega,\varepsilon_{M}\right)=-e^{2}\int\frac{d^{2}k}{\left(2\pi\right)^{2}}\varepsilon_{\alpha\beta}\Omega\left(\mathbf{k}\right)f_{0T}\left(\varepsilon_{M}\left(\mathbf{k}\left(t\right)\right)\right)-e\int\frac{d^{2}k}{\left(2\pi\right)^{2}}\nabla_{{\bf k_{\alpha}}}\varepsilon_{M}\left({\bf k}\right)\frac{\partial f_{0T}\left(\varepsilon_{M}\left(\mathbf{k}_{0}(t)\right)\right)}{\partial\varepsilon}\times\nonumber \\
 & \times\int_{-\infty}^{t}dt'D^{-1}\left(\mathbf{k}_{0}\left(t\right)\right)\nabla_{\mathbf{k_{\beta}}}\varepsilon_{M}\left(\mathbf{k}_{0}\left(t'\right)\right)\cdot\exp\left(i\omega\left(t'-t\right)\right)\exp\left(-\left(t-t'\right)/\tau_{S}\right)
\label{eq:Topological_regular}
\end{align}

with 
$\sigma_{\alpha\beta}^{top}\left(e,\tau_{S},T,\mathbf{B},\omega,\varepsilon_{M}\right)=-e^{2}\int\frac{d^{2}k}{\left(2\pi\right)^{2}}\varepsilon_{\alpha\beta}\Omega\left(\mathbf{k}\right)f_{0T}\left(\varepsilon_{M}\left(\mathbf{k}\left(t\right)\right)\right)$
and 
\begin{align}
\sigma_{\alpha\beta}^{reg}\left(e,\tau_{S},T,\mathbf{B},\omega,\varepsilon_{M}\right) & =-e\int\frac{d^{2}k}{\left(2\pi\right)^{2}}\nabla_{{k_{\alpha}}}\varepsilon_{M}\left({\bf k}\right)\frac{\partial f_{0T}\left(\varepsilon_{M}\left(\mathbf{k}_{0}(t)\right)\right)}{\partial\varepsilon}\times\nonumber \\
 & \int_{-\infty}^{t}dt'D^{-1}\left(\mathbf{k}_{0}\left(t'\right)\right)\nabla_{{k_{\beta}}}\varepsilon_{M}\left(\mathbf{k}_{0}\left(t'\right)\right)\cdot\exp\left(i\omega\left(t'-t\right)\right)\exp\left(-\left(t-t'\right)/\tau_{S}\right)
\end{align}

Where we assume a constant relaxation time ${\tau}_{S}$. By introducing
$\mathcal{B}=eB$, $\tilde{t}=\left(t'-t\right)/\tau_{S}$, $\mathcal{E}_{M}=\tau_{S}\varepsilon_{M},\,\tilde{\omega}=\omega\tau_{S}$ and taking into account the equations of motion,
we obtain for the regular part of the conductivity: 
\begin{align}
 & \sigma_{\alpha\beta}^{reg}\left(e,\tau_{S},T,\mathbf{B},\omega,\varepsilon_{M}\right)=-e\tau_{S}^{-1}\int\frac{d^{2}k}{\left(2\pi\right)^{2}}\nabla_{{\bf k_{\alpha}}}\mathcal{E}_{M}\left({\bf k}\right)\frac{\partial f_{0T\tau_{S}}\left(\mathcal{E}_{M}\left(\mathbf{k}\right)\right)}{\partial\mathcal{E}}\times\nonumber \\
 & \times\int_{-\infty}^{0}D^{-1}\left(\mathbf{k}_{0}\left({\tilde{t}}\right)\right)\nabla_{k_{\beta}}\mathbf{\mathcal{E}}_{M}\left(\mathbf{k}_{0}\left({\tilde{t}}\right)\right)\cdot\exp\left(i\tilde{\omega}{\tilde{t}}\right)\exp\left({\tilde{t}}\right)d{\tilde{t}}=e\tau_{S}^{-1}\sigma_{\alpha\beta}^{reg}\left(1,1,T\tau_{S},e\mathbf{B},\omega\tau_{S},\tau_{S}\varepsilon_{M}\right)\label{eq:Scaling-3}
\end{align}
where in the above equation, we have used the notation:
\begin{equation}
\frac{\partial f_{0T}\left(\varepsilon_{M}\left(\mathbf{k}\right)\right)}{\partial\varepsilon} \equiv \frac{\partial}{\partial\varepsilon_{M}}\left(\frac{1}{1+\exp\left(\beta\varepsilon_{M}\right)}\right) = \frac{\partial}{\partial\mathcal{E}_{M}}\left(\frac{1}{1+\exp\left(\beta\mathcal{E}_{M}/\tau_{S}\right)}\right) \equiv \frac{\partial f_{0T\tau_{S}}\left(\mathcal{E}_{M}\left(\mathbf{k}\right)\right)}{\partial\mathcal{E}}
\end{equation}

\end{widetext}

This scaling relation will be useful for the numerical calculations
in the following sections. In addition, as $\sigma_{\alpha\beta}^{reg}\left(1,1,T\tau_{S},e\mathbf{B},\omega\tau_{S},\tau_{S}\varepsilon_{M}\right)$
is a power series in $eB$, it is impossible to completely disentangle
all the powers of $e$ that enter the expression for the conductivity.

\section{\label{sec:Application:-regular-van}Lifshitz transition of neck formation/collapse on rectangular
lattice}

We use the frequency dependent Chambers formula to calculate the
components of the conductivity tensor in the case of an energy dispersion
of the form

\begin{equation}
\varepsilon\left(k\right)=-2t_{x}\cos\left(k_{x}\right)-2t_{y}\cos\left(k_{y}\right)\;\left(t_{y}>t_{x}\right)
\end{equation}
We define $\mu_{c}=2\left(t_{y}-t_{x}\right)$ and $\mu_{0}=2\left(t_{y}+t_{x}\right)$.
For this band we've got the following topologies:

\begin{align}
-\mu_{0}<\mu<-\mu_{c} & \qquad electron\,pockets\nonumber \\
-\mu_{c}<\mu<\mu_{c} & \qquad open\,Fermi\,surface\nonumber \\
\mu_{c}<\mu<\mu_{0} & \qquad hole\,pockets\label{eq:vaR_Hove}
\end{align}
The components of $\sigma$ for all three cases read (the details
of all calculations in this section are left in Appendix \ref{sec:Hall-coeffcient-calculations}):

\begin{equation}
\sigma_{xx}^{i}=\frac{2\sigma_{0}}{K}\sum_{n}\frac{\left[1+i\omega\tau_{S}\right]sech^{2}\left[\frac{n\pi K^{'}}{2K}\right]\sin^{2}\left[\frac{n\pi u_{i}}{2K}\right]}{\left[1+i\omega\tau_{S}\right]^{2}+\left[n\omega_{c}\tau_{S}\right]^{2}}\label{eq:sigma_xx}
\end{equation}

\begin{align}
\sigma_{yy}^{i} & =\frac{\sigma_{0}\delta_{i,o}}{K}\frac{1}{1+i\omega\tau_{S}}+\nonumber \\
 & +\frac{2\sigma_{0}}{K}\sum_{n}\frac{\left[1+i\omega\tau_{S}\right]sech^{2}\left[\frac{n\pi K^{'}}{2K}\right]\cos^{2}\left[\frac{n\pi u_{i}}{2K}\right]}{\left[1+i\omega\tau_{S}\right]^{2}+\left[n\omega_{c}\tau_{S}\right]^{2}}\label{eq:Sigma_yy}
\end{align}

\begin{align}
\sigma_{xy}^{i} & =\left(\delta_{i,o}+\delta_{i,e}-\delta_{i,h}\right)\frac{\sigma_{0}}{K}\times\nonumber \\
 & \times\sum_{n}\frac{\left(n\omega_{c}\tau_{S}\right)sech^{2}\left[\frac{n\pi K^{'}}{2K}\right]\sin\left[\frac{n\pi u_{i}}{K}\right]}{\left[1+i\omega\tau_{S}\right]^{2}+\left[n\omega_{c}\tau_{S}\right]^{2}}\label{eq:sigma_xy}
\end{align}
where $h$ stands for hole orbits, $e$ stands for electron orbits
and $o$ stands for open orbits. Furthermore, we have defined that
$\kappa=\sqrt{\frac{\mu_{0}^{2}-\mu^{2}}{\mu_{0}^{2}-\mu_{c}^{2}}}$,
$\omega_{0}=eB\sqrt{4t_{x}t_{y}}$, $m_{0}=\frac{1}{\sqrt{4t_{x}t_{y}}}$,
$\sigma_{0}=e^{2}\tau_{S}\sqrt{4{t_{x}}{t_{y}}}$ and $K\left(\kappa\right)$
is the complete elliptic integral of the first kind. In the case of
closed surfaces $K\equiv K\left(\kappa\right),\,K^{'}\equiv K\left(\sqrt{1-\kappa^{2}}\right)$.
For open surfaces we substitute $K\left(\kappa\right)\rightarrow\frac{1}{\kappa}K\left(1/\kappa\right)$
and similarly for $K^{'}$ \cite{Maharaj_2017}. In addition

\begin{equation}
\omega_{c}=\begin{cases}
\frac{\pi\omega_{0}}{2K\left(\kappa\right)} & closed\:orbits\\
\frac{\pi\kappa\omega_{0}}{2K\left(1/\kappa\right)} & open\:trajectories
\end{cases}
\end{equation}
Finally $u_{i}$ and $u_{o}$ are defined via Jacobean elliptic functions

\begin{eqnarray}
sn\left(u_{e},\kappa\right)=\sqrt{\frac{\mu_{0}-\mu_{c}}{\mu_{0}-\mu}}\\
sn\left(u_{h},\kappa\right)=\sqrt{\frac{\mu_{0}-\mu_{c}}{\mu_{0}+\mu}}\\
sn\left(\kappa u_{o},1/\kappa\right)=\sqrt{\frac{\mu_{0}+\mu}{\mu_{0}+\mu_{c}}}
\end{eqnarray}
For closed Fermi surfaces the sums are over positive odd integers
while for open surfaces the sums are over positive even integers.

\begin{figure}[!]
\begin{minipage}[t]{0.9\columnwidth}
  \includegraphics[width=8cm]{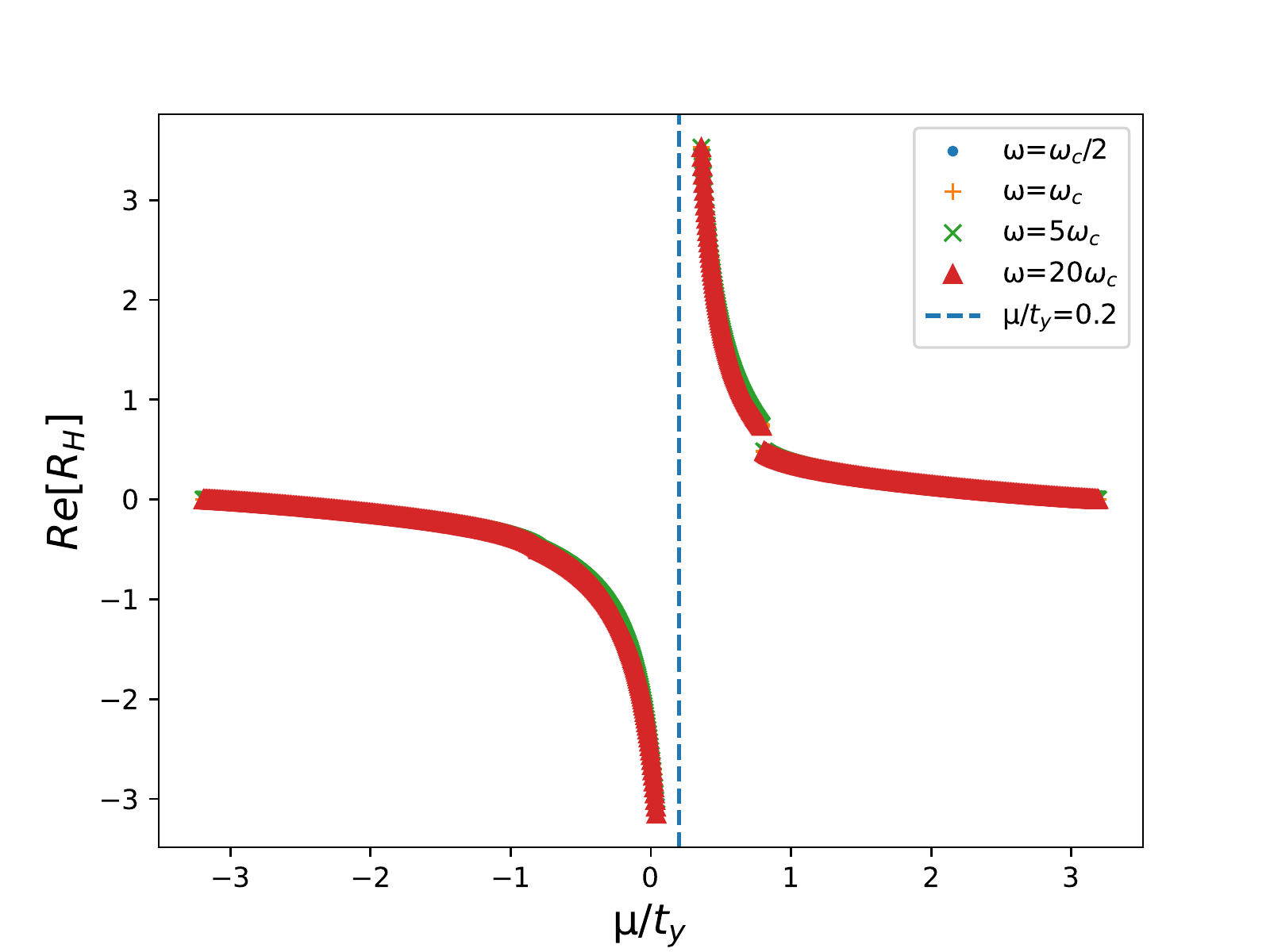} 
\end{minipage}\hfill{} 
\begin{minipage}[t]{0.9\columnwidth}
 \includegraphics[width=8cm]{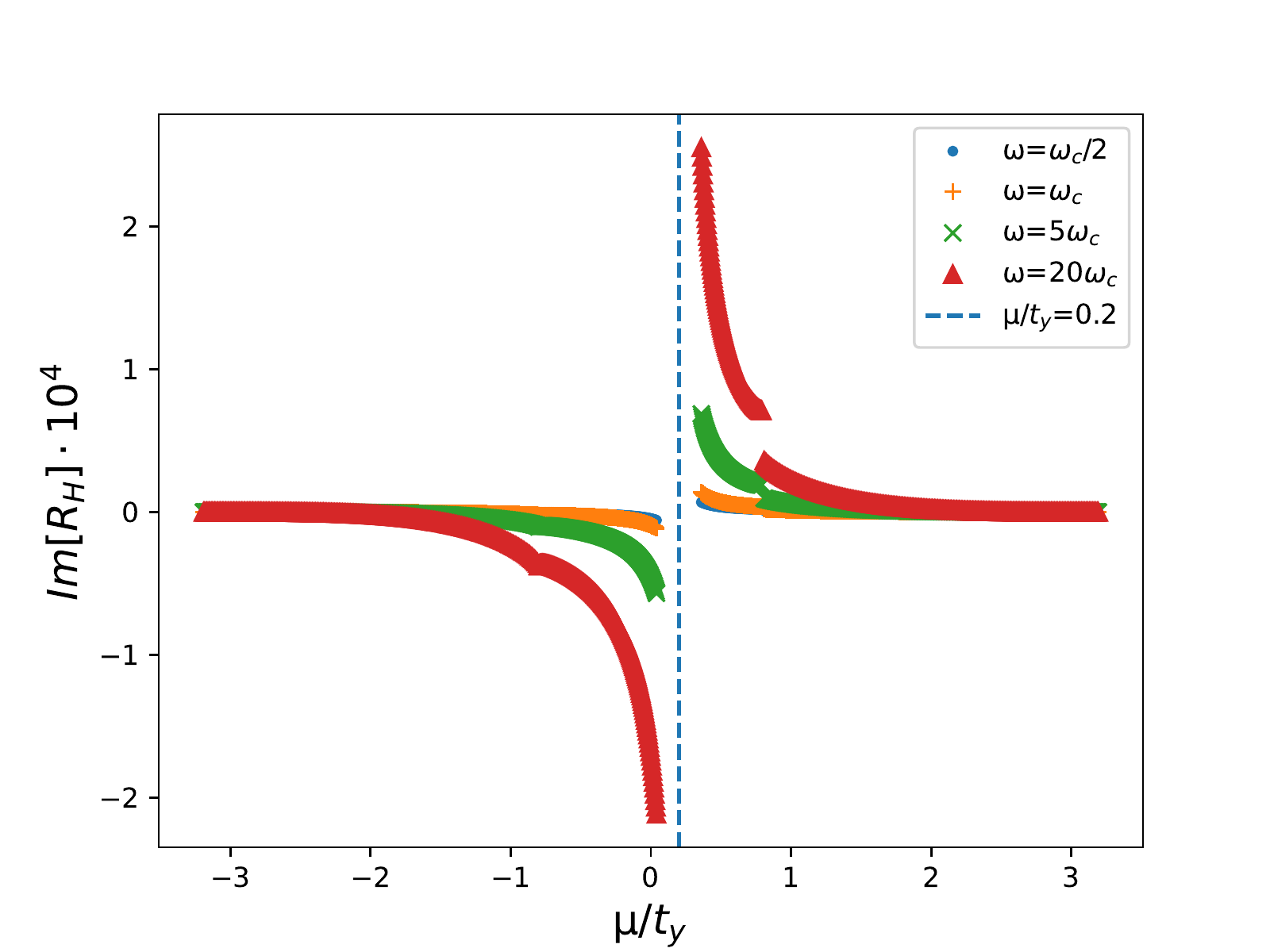} 
\end{minipage}\caption{\label{fig:Rectangular_R_Hall}The real and imaginary part of the
Hall number for four different frequencies, B=0.01 and $\omega_{c}\tau_{S}=0.01$ for the example of the rectangular lattice.
The Lifshitz transition occurs is at $\mu=0.25$. The results are valid away form $\mu=0.25$ where magnetic breakdown phenomena should be taken into account for a full analysis.}
\end{figure}

\begin{figure}[!]
\begin{minipage}[t]{0.9\columnwidth}
  \includegraphics[width=8cm]{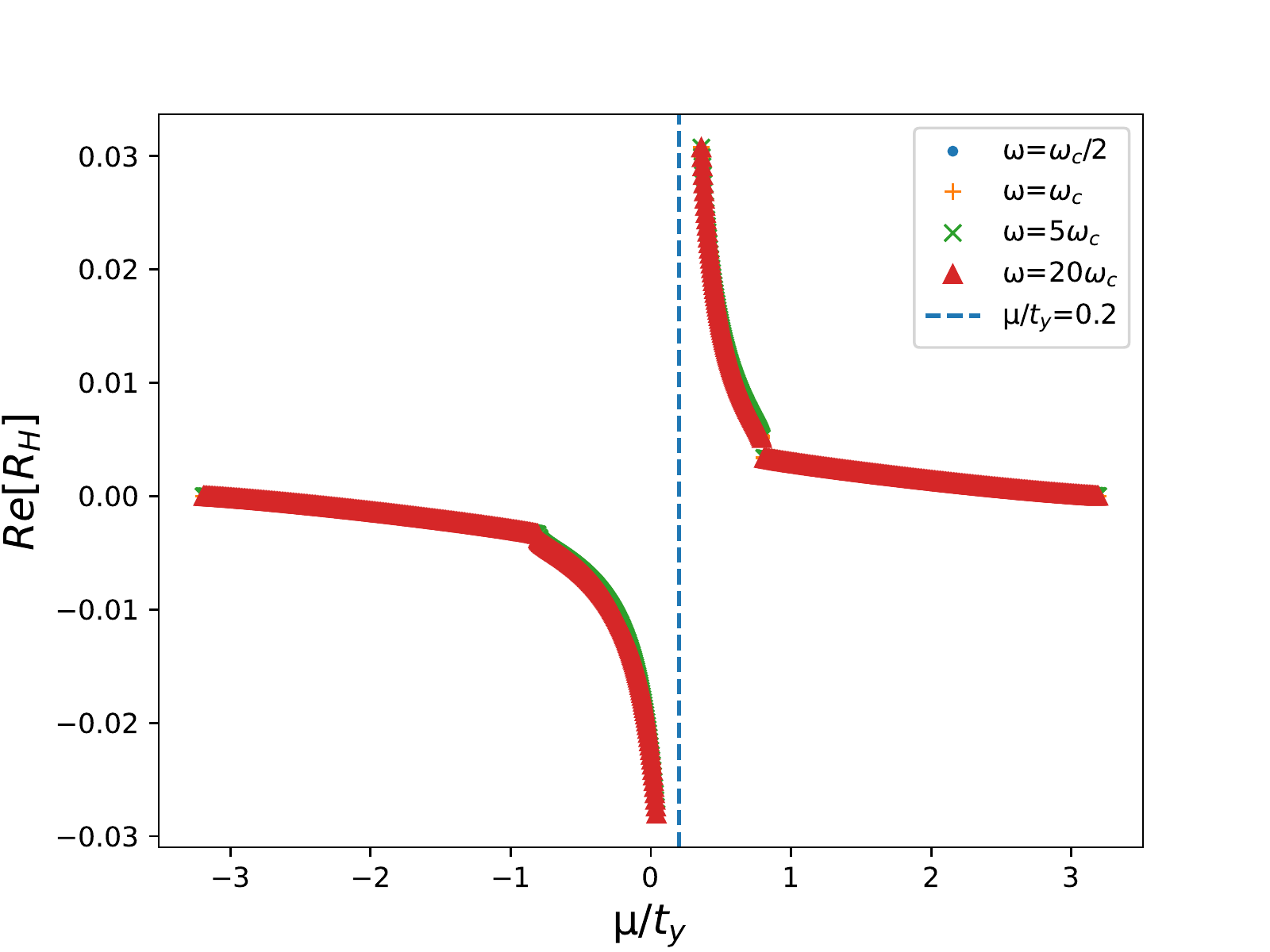} 
\end{minipage}\hfill{} 
\begin{minipage}[t]{0.9\columnwidth}
\includegraphics[width=8cm]{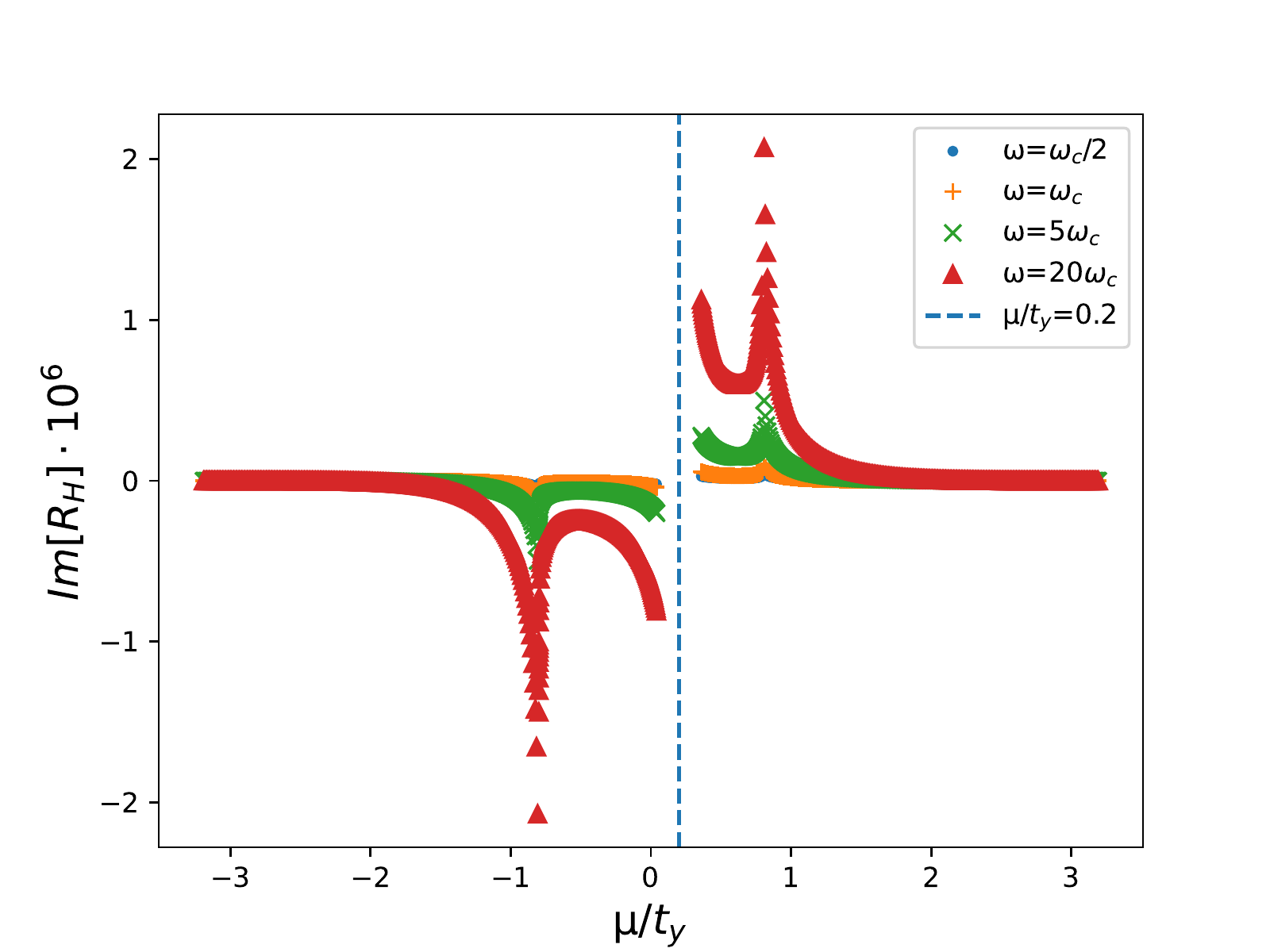} 
\end{minipage}\caption{\label{fig:Rectangular_R_Hall} Same as in Fig. 1 the real and imaginary
part of the Hall number for four different frequencies and now for
$\omega_{c}\tau_{S}=1$. As before, the results are valid away from the Lifshitz transition.}
\end{figure}

The calculation of the Hall number for both high and low magnetic
fields leads to expressions that are independent of the frequency
and similar to the ones calculated in \cite{Maharaj_2017}. For completeness
we present the formulae in the Appendix C1. In Figs (1) and (2) we
present the results for two limiting values of $\omega_{c}\tau_{S}=0.01\ll1$
and for $\approx1$. For each of these values several different frequencies
have been chosen. The magnetic field is $B=0.01$ in units of $hc/(a^{2}e)$
where $a$ is the lattice constant (for simplicity we work with $a=h=c=1$).
Consistently with the fact that at $\mu/t_{y}=0.25$, there is a regular
(logarithmic) Van Hove singularity (in the language of Lifshitz transitions
is classified as a neck formation/collapse), the real part of the
Hall coefficient changes sign from electron to hole-like. At values
of $\mu/t_{y}$ close to $\pm1$ the discontinuities signal the change
from closed pockets to open Fermi surface. The real part do not show
any significant frequency dependence. On the contrary, the imaginary
part displays a strong frequency dependence with all features (change
of sign at the Lifshitz transition and discontinuity at close-open
Fermi surface transition) becoming more pronounced with increasing
$\omega$. Away from these special values of $\mu/t_{y}$, the imaginary
part of the Hall number is practically zero, reflecting that the dispersion
relation is parabolic. Close to the Lifshitz transition quantum tunnelling must be taken into account for full quantitative analysis \cite{Falicov,Glazman}.

\section{\label{subsec:Application:-supermetal-saddle}Supermetal: a simple
model for highly doped graphene }

\subsection{\label{subsec:Tight-binding-models}A relevant tight binding model}

We will consider a Hamiltonian on hexagonal lattice which is relevant
to recent studies of graphene that showed the emergence of a higher
order Van Hove singularity \cite{Yuan_Isobe_Fu,Isobe_Fu}. By considering
third nearest neighbors tunnelling the simple Hamiltonian that
generates higher order Van Hove singularities, as a consequence of FSTTs, reads:

\begin{equation}
H=\left(\begin{array}{cc}
0 & f\left({\bf k}\right)\\
f^{*}\left({\bf k}\right) & 0
\end{array}\right)\label{eq:Hamiltonian-2}
\end{equation}
Where 
\begin{align}
f\left({\bf k}\right) & =-t\left[e^{-ik_{y}}+e^{\frac{i}{2}\left(\sqrt{3}k_{x}+k_{y}\right)}+e^{\frac{i}{2}\left(-\sqrt{3}k_{x}+k_{y}\right)}\right]\nonumber \\
 & -c\left[e^{i2k_{y}}+e^{-i\left(\sqrt{3}k_{x}+k_{y}\right)}+e^{-i\left(-\sqrt{3}k_{x}+k_{y}\right)}\right]\label{eq:Dispersion-1}
\end{align}
and energy dispersion $E_{\pm}\left(k\right)=\pm\sqrt{f^{*}\left({\bf k}\right)f\left({\bf k}\right)}$.
\begin{figure}[!]
\begin{minipage}[t]{0.45\columnwidth}
\includegraphics[width=4cm]{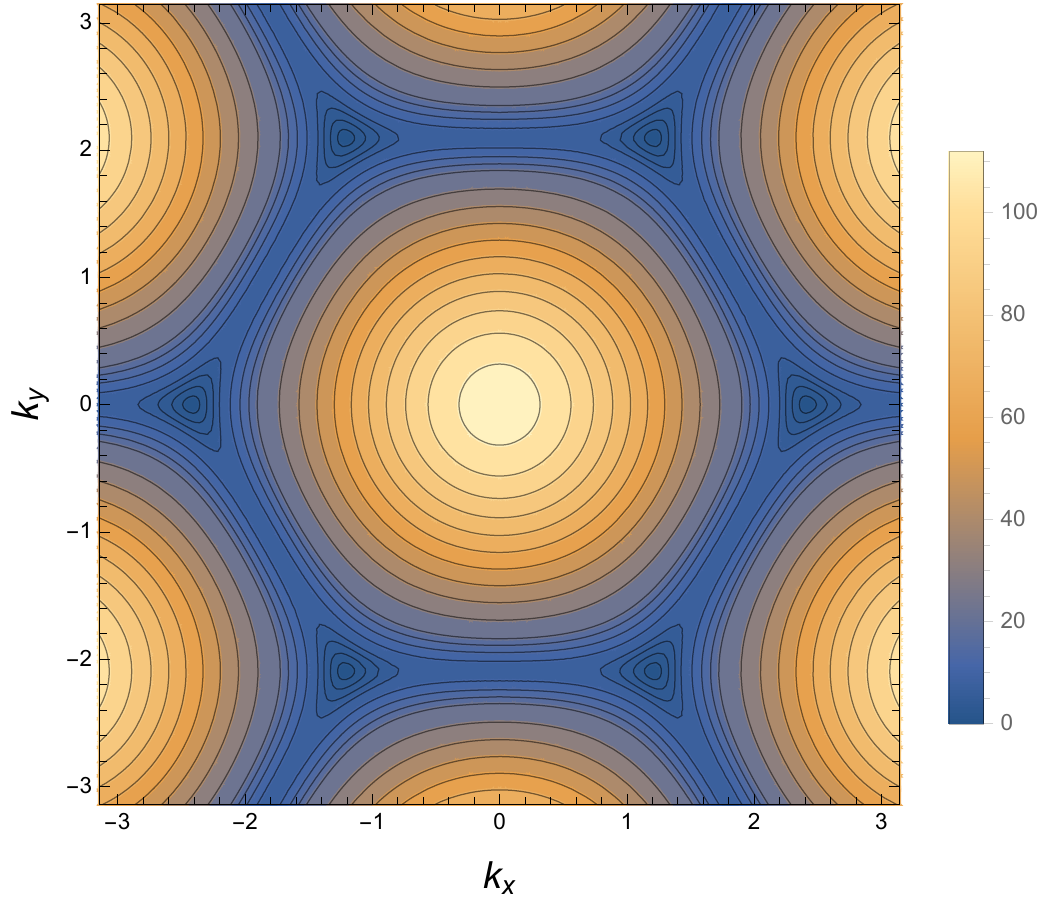} 
\end{minipage}\hfill{} 
\begin{minipage}[t]{0.45\columnwidth}
\includegraphics[width=4cm]{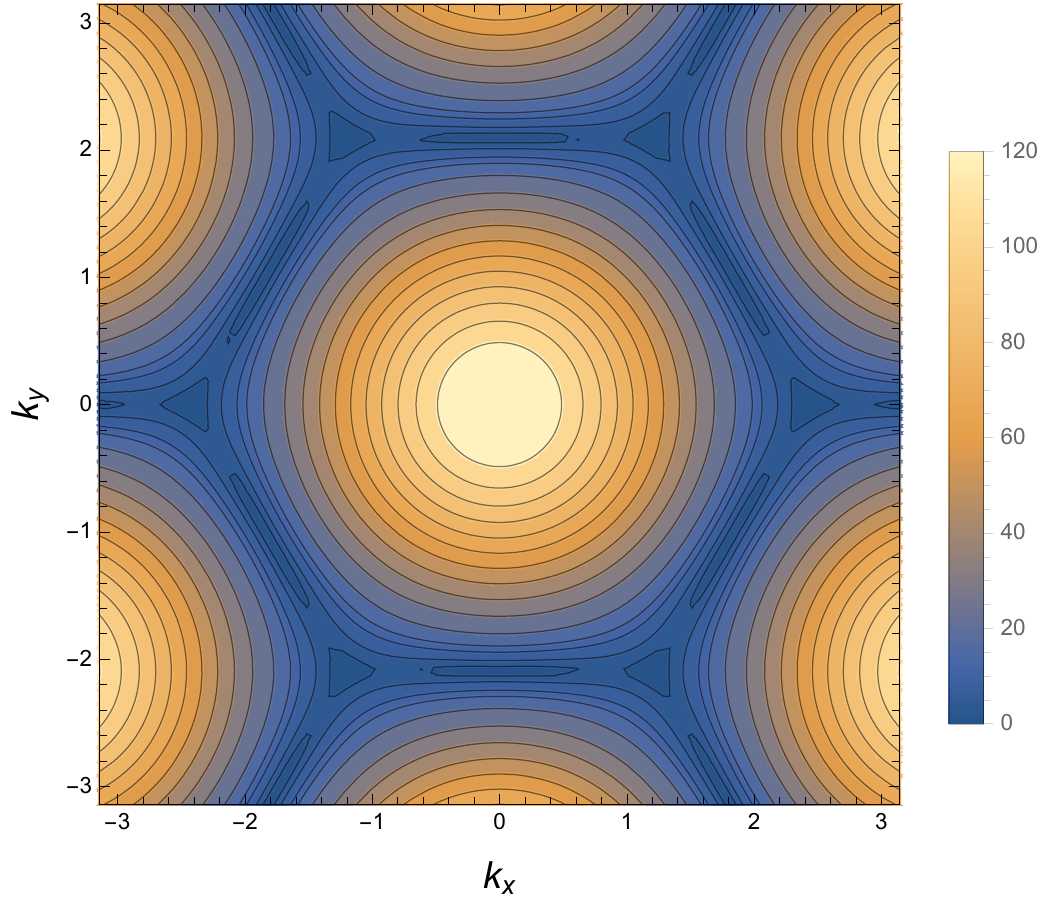} 
\end{minipage}\caption{Dispersion for graphene with third
nearest neighbors for $c/t=1/4$ where the higher order Van Hove singularity
(cusp, on the left) can be visualized at ${\bf G}=\frac{2\pi}{3}(0,1)$.
For $c/t=1/3$ (on the right) it is a nodal point with ill-defined
gradient. For the plots we take $t=30$) .}
\label{fig:Graphene_third_nearest}
\end{figure}

This leads to a higher order Van Hove saddle in both $E_{\pm}\left({\bf k}\right)$
at wave vector ${\bf G}=\frac{2\pi}{3}(0,1)$ (for lattice constant
$a=1$) for $c=\frac{t}{4}$ while for $c=\frac{t}{3}$ it is a nodal
point with closure of gap and ill defined gradient. It is rather instructive
to present the contour plots of $E_{+}$ for both cases in Figure
(\ref{fig:Graphene_third_nearest}). The higher order saddle for $c=\frac{t}{4}$
is a cusp at ${\bf G}=\frac{2\pi}{3}(0,1)$, based on the catastrophe
theory classification \cite{Chandrasekaran,LiangFu}.

\begin{figure}[!]
\centering
\includegraphics[width=0.4\textwidth]{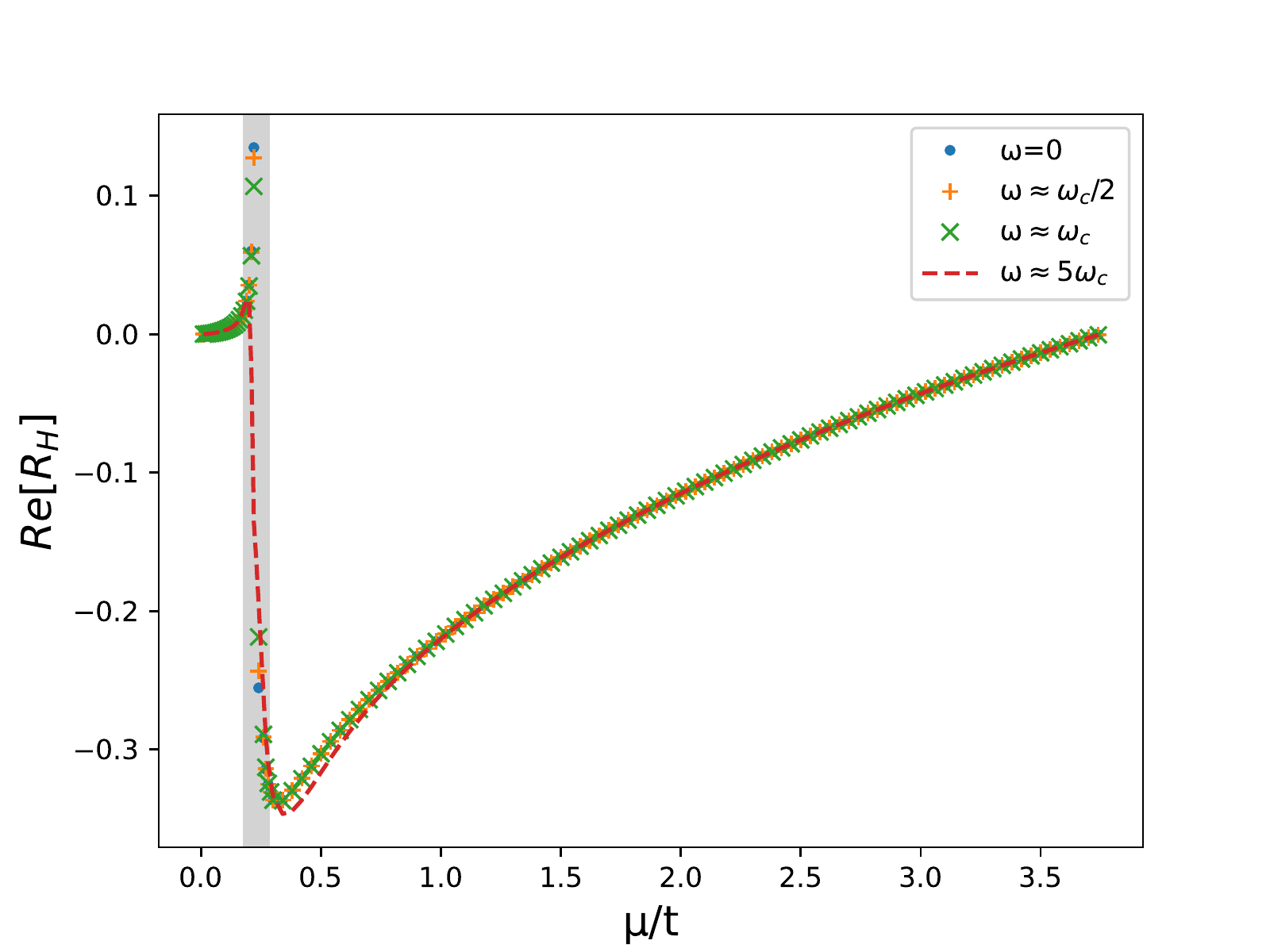} 
\includegraphics[width=0.4\textwidth]{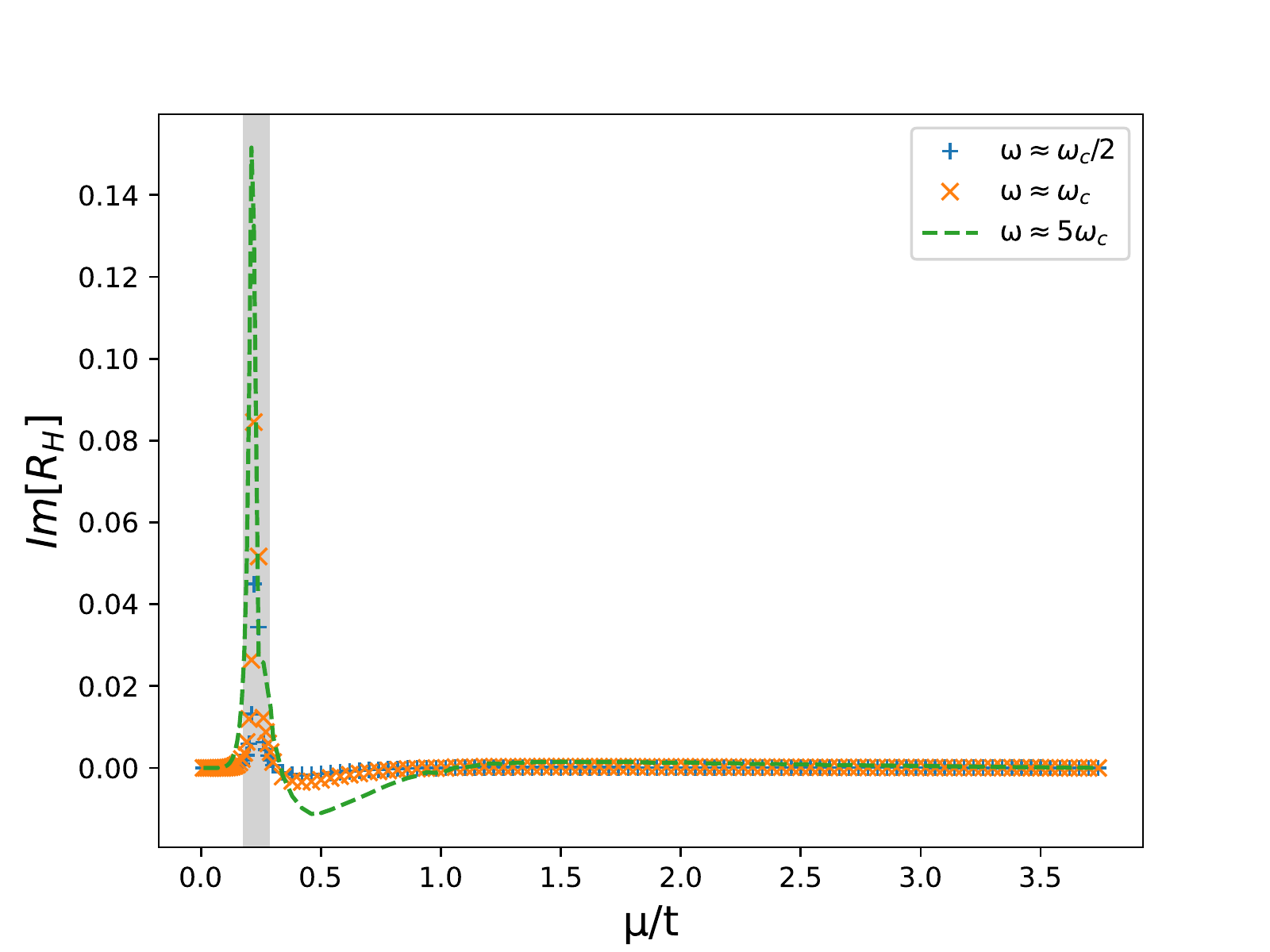} 
\caption{\label{fig:Graphene_R_Hall} Real and imaginary part of Hall coefficient
for $\omega_{c}\tau_{S}=0.3$.}
\includegraphics[width=0.4\textwidth]{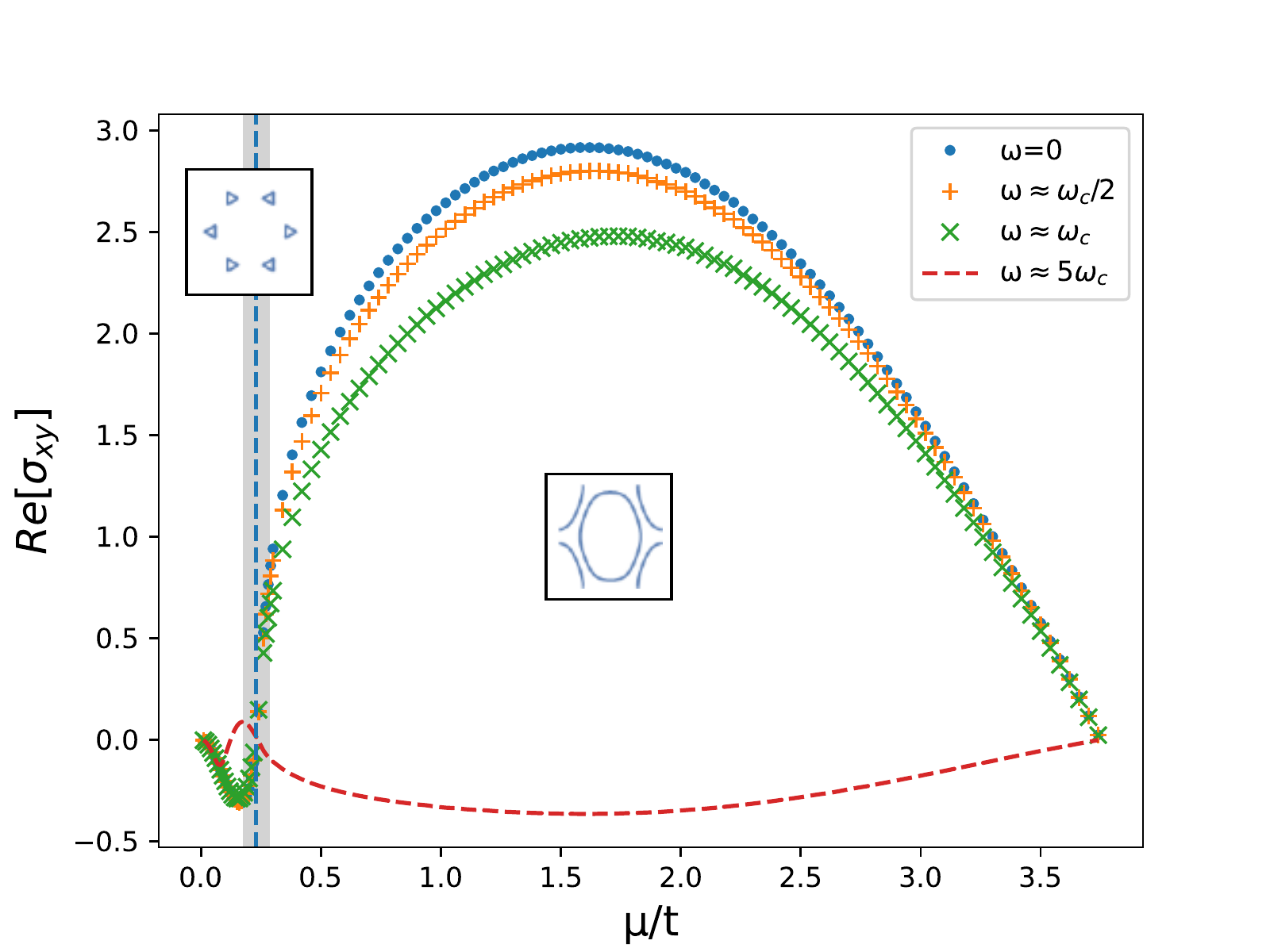} 
\includegraphics[width=0.4\textwidth]{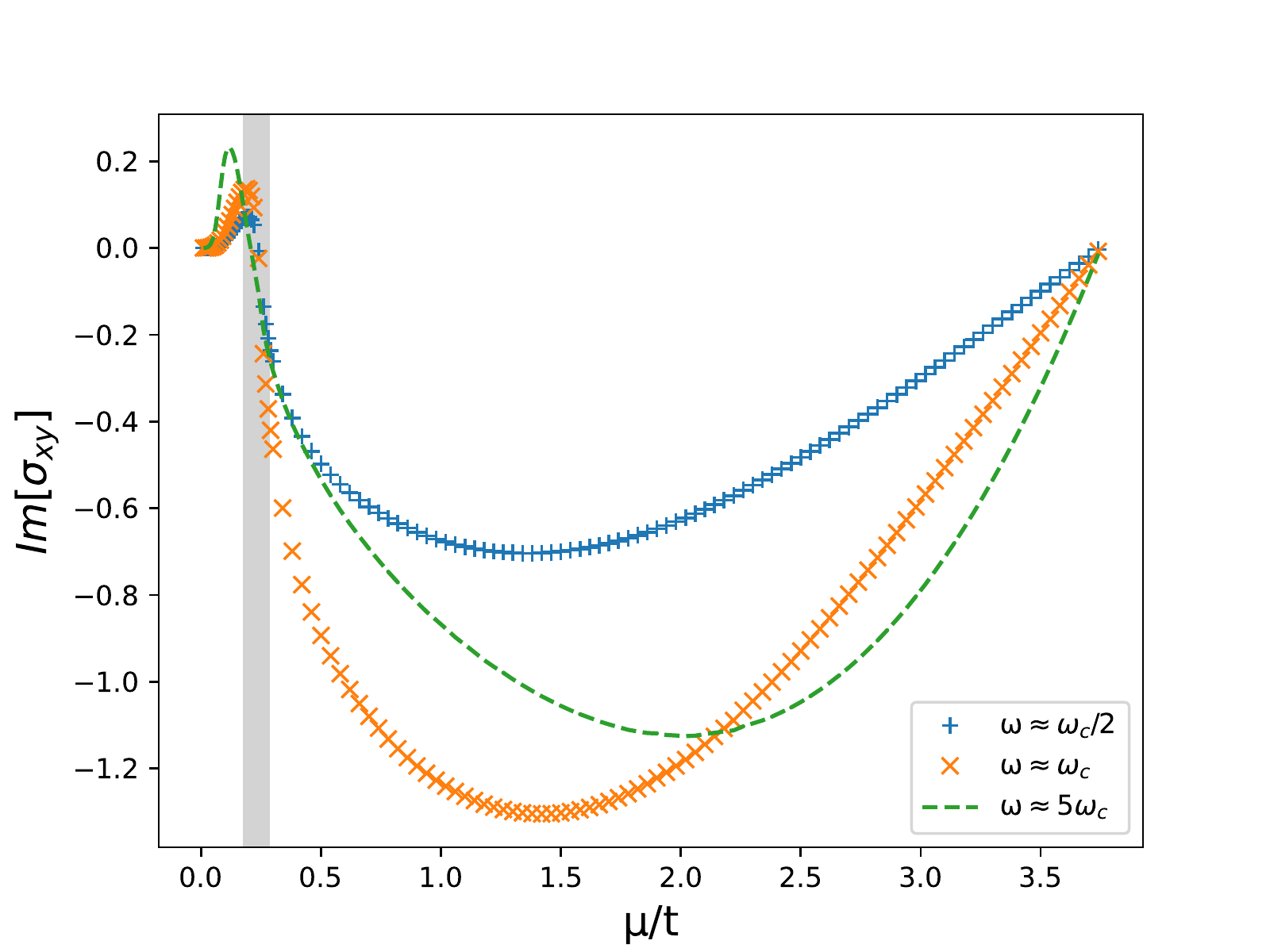} 
\caption{Real and imaginary part of Hall conductivity for $\omega_{c}\tau_{S}=0.3$ for the simple model of highly doped graphene.
On the left figure, the Fermi surfaces in different regimes have been
depicted, so the topological transition is evident. The shaded areas, around FSTTs, denote where magnetic breakdown phenomena should be taken into account for a full quantitative analysis}
\label{fig:Graphene_sigma_xy_tausmall} 
\end{figure}

\begin{figure}[!]
\includegraphics[width=0.4\textwidth]{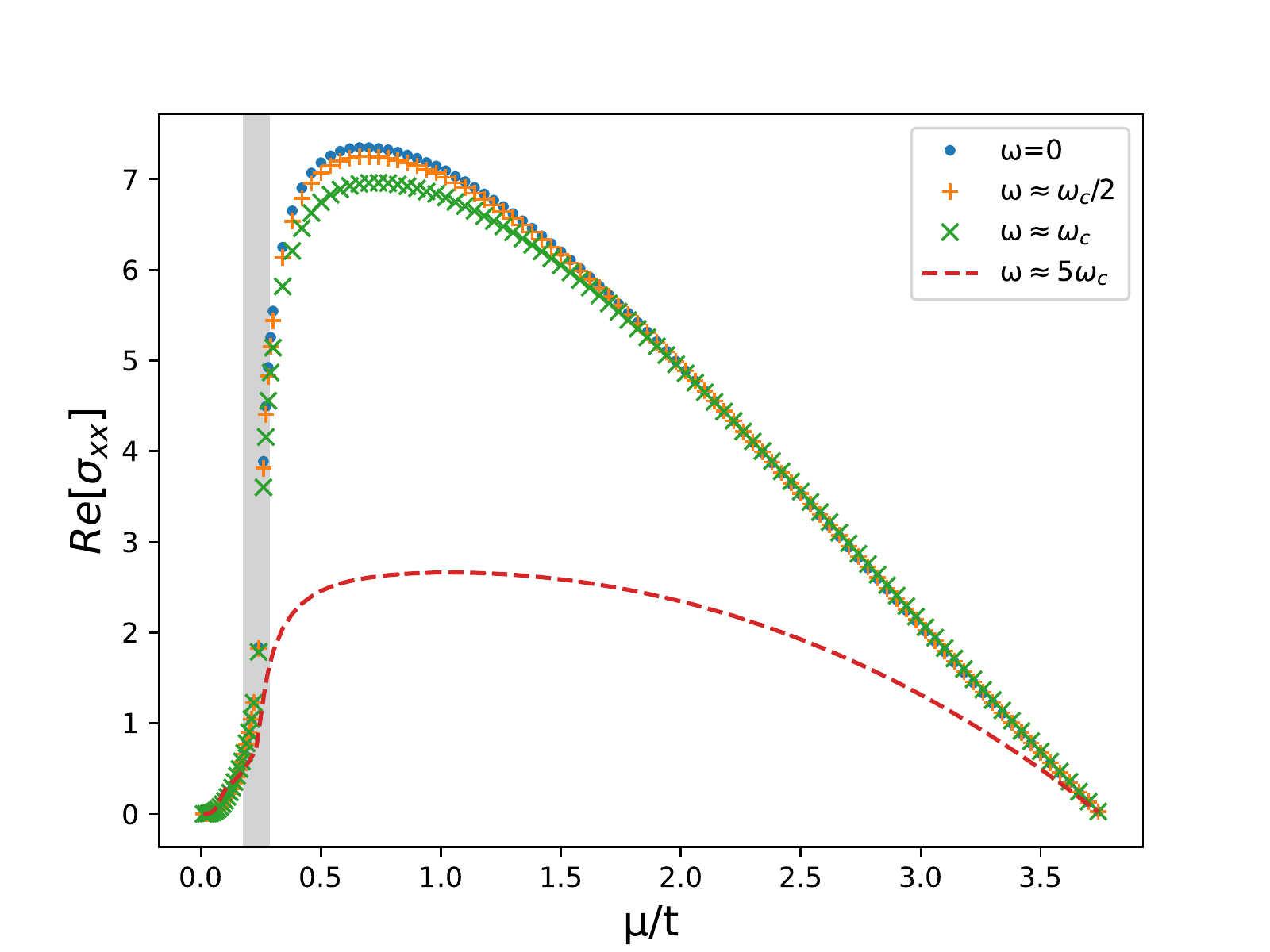} 
\includegraphics[width=0.4\textwidth]{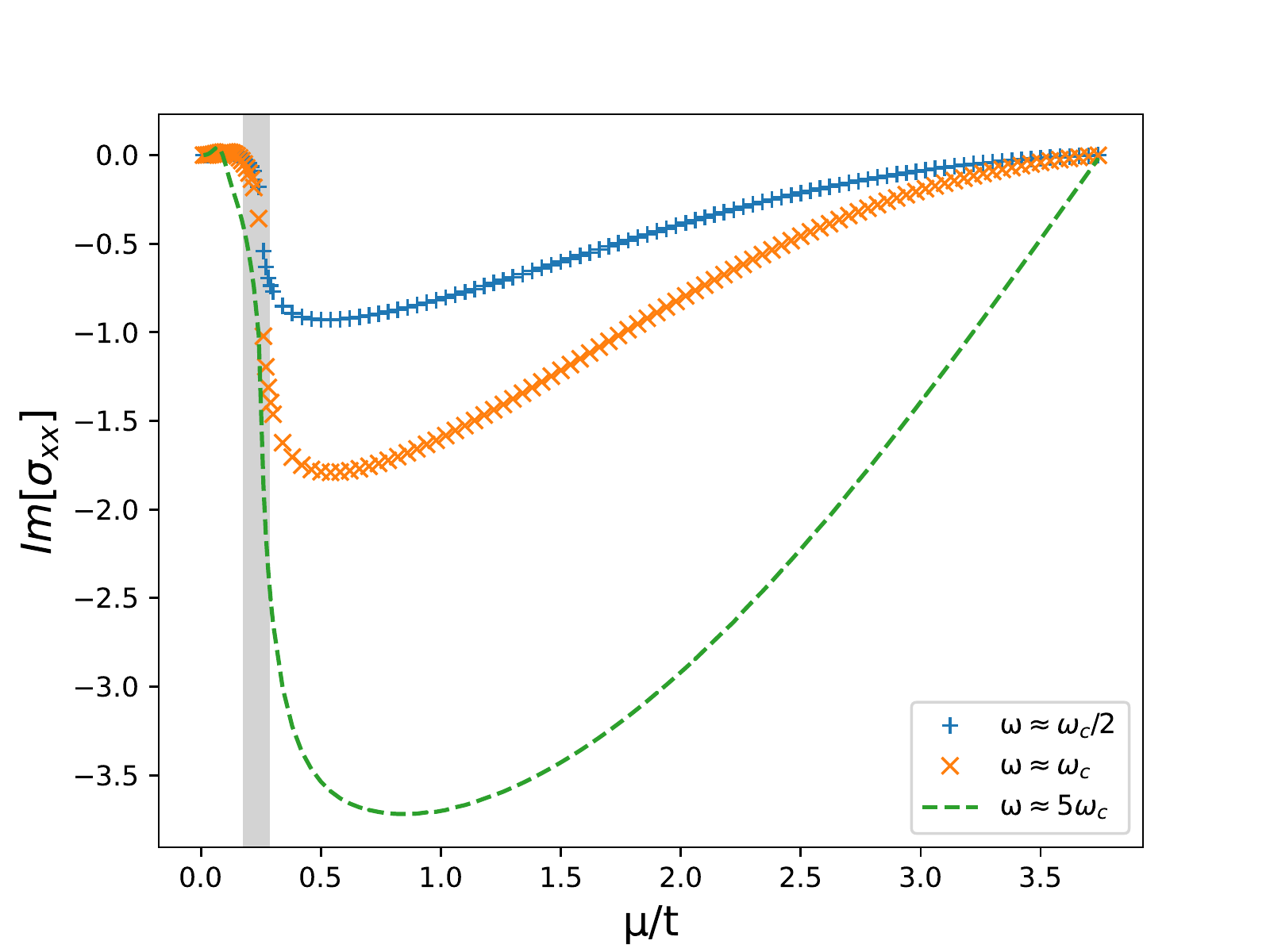} 
 \caption{Real and imaginary part of the conductivity $\sigma_{xx}$ (which
is the same as $\sigma_{yy}$) for $\omega_{c}\tau_{S}=0.3$.}
\label{fig:Graphene_sigma_xx_yy_tausmall} 
\end{figure}

\subsection{\label{subsec:Numerical-simulations-1}Numerical results}

\begin{figure}[!]
\centering
\includegraphics[width=0.4\textwidth]{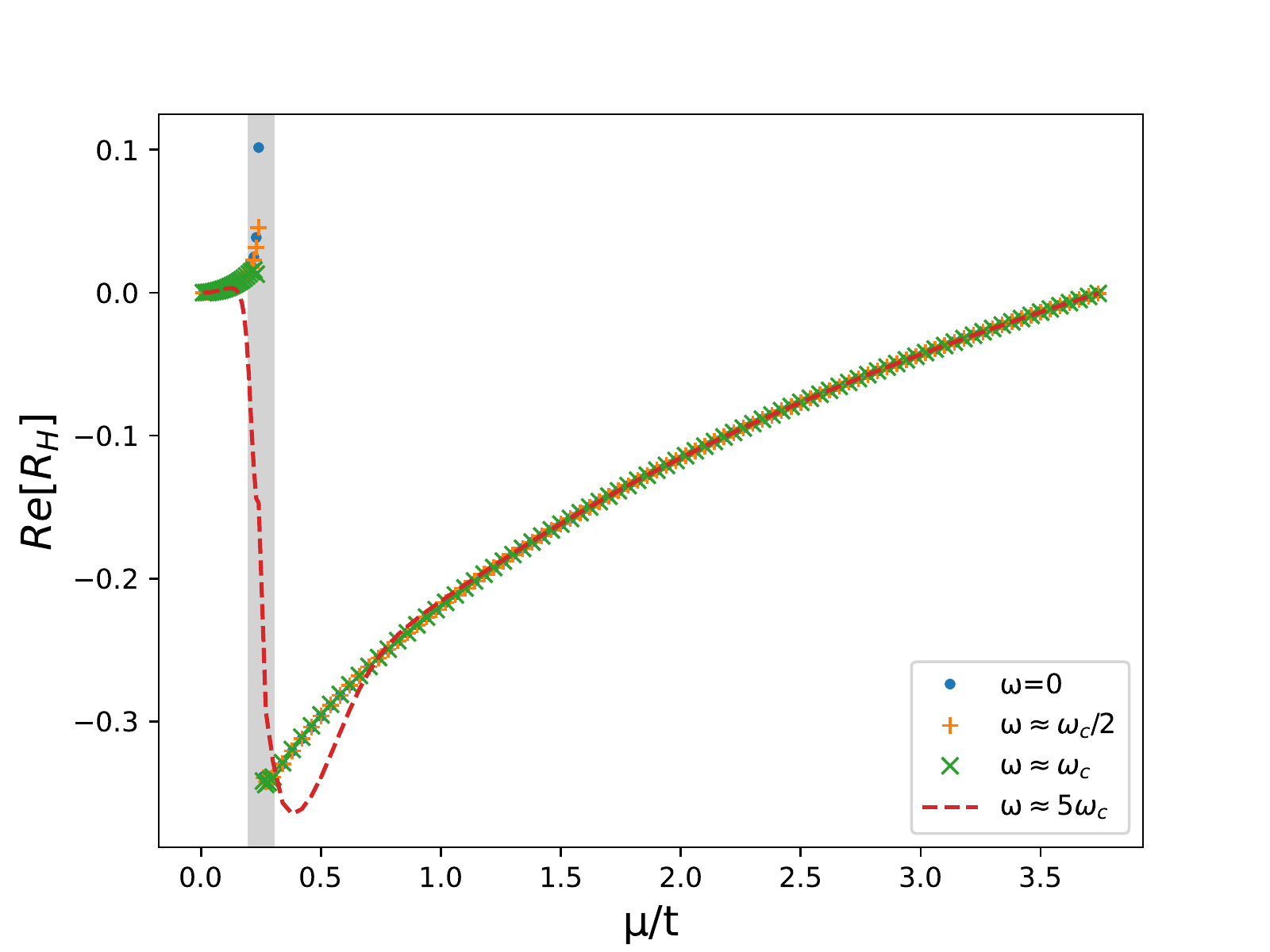}
\includegraphics[width=0.4\textwidth]{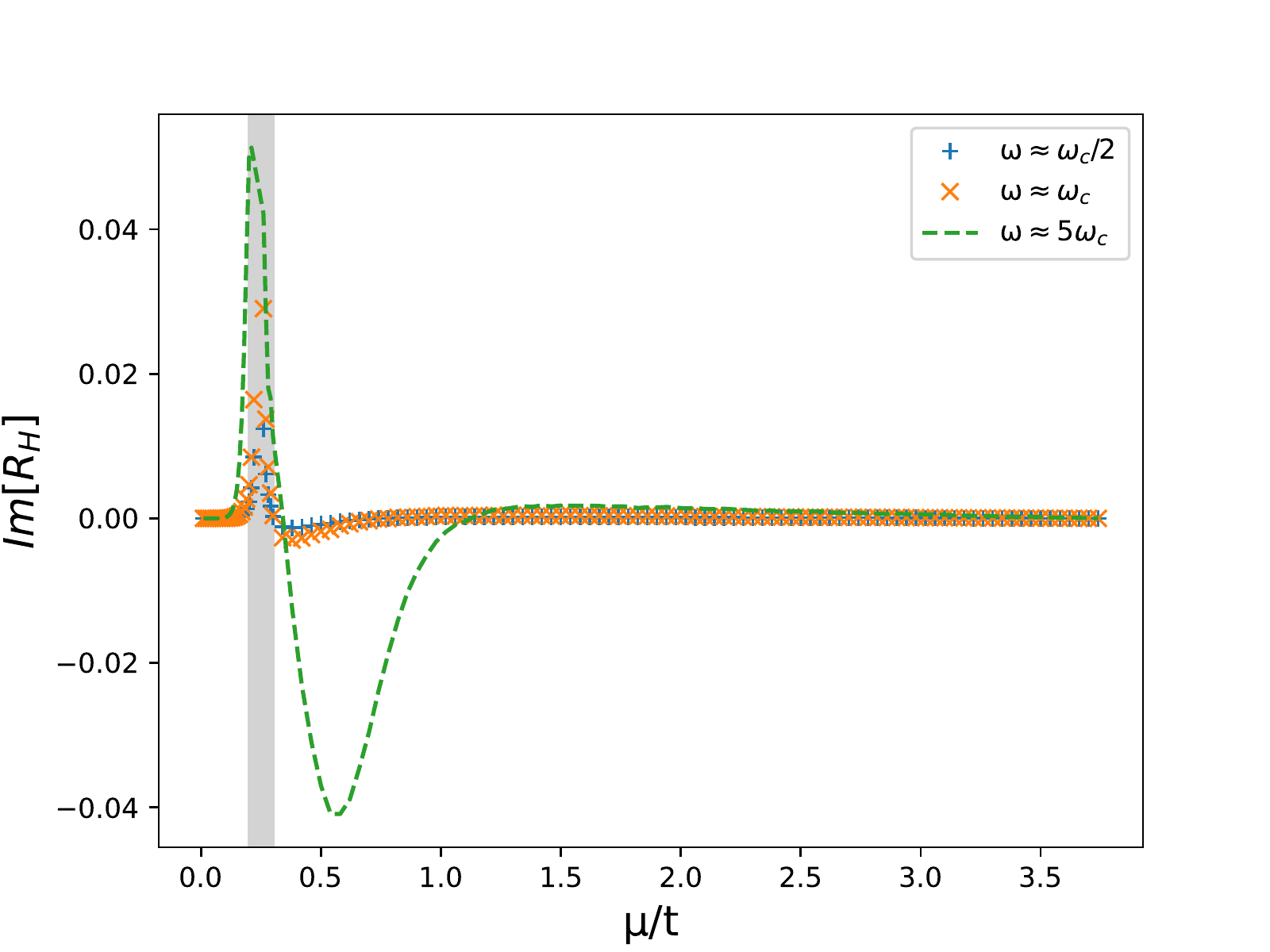}
\caption{Real and imaginary part of Hall coefficient for $\omega_{c}\tau_{S}=1$.}
\label{fig:Graphene_R_Hall} 
\includegraphics[width=0.4\textwidth]{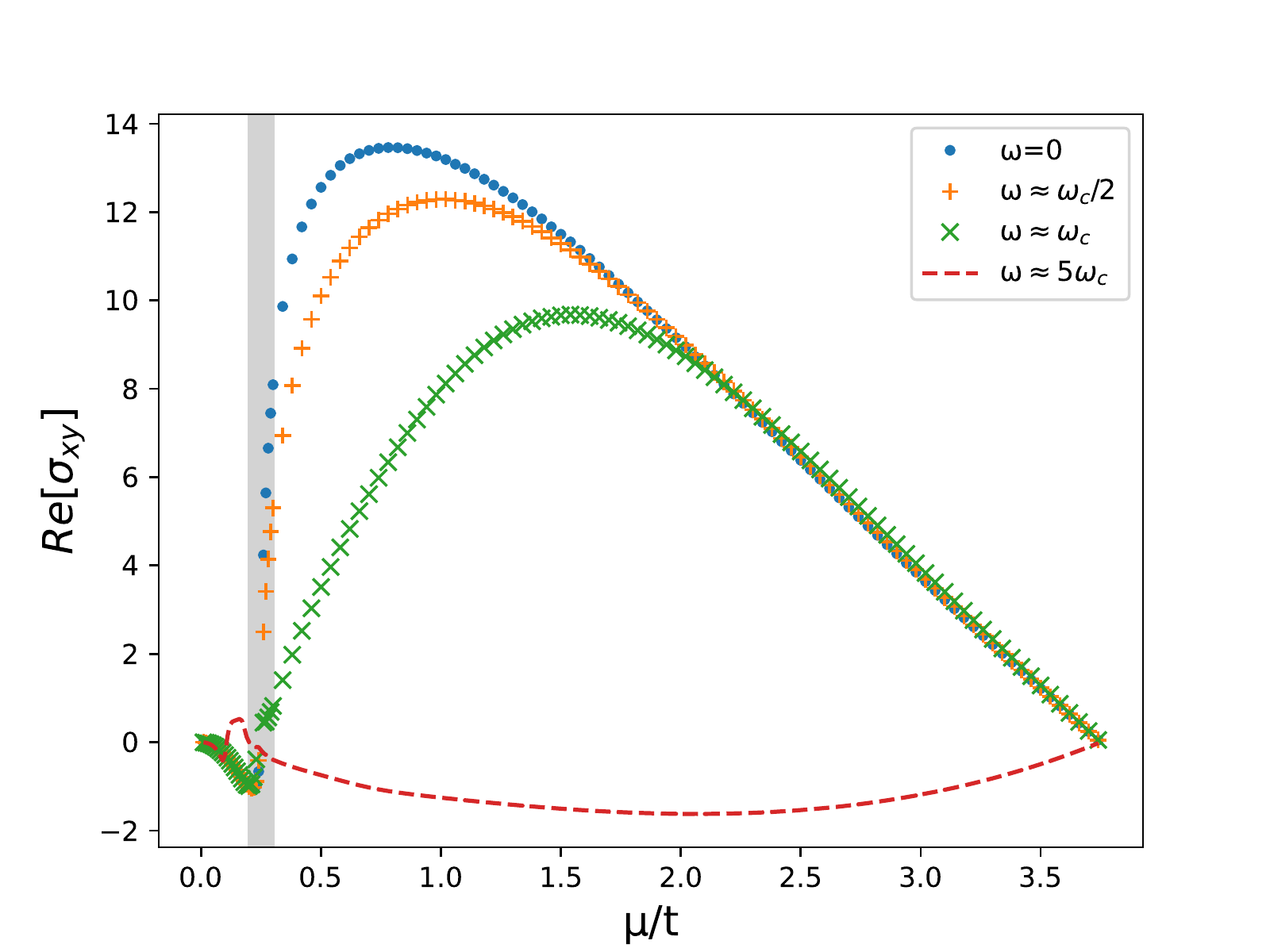}
\includegraphics[width=0.4\textwidth]{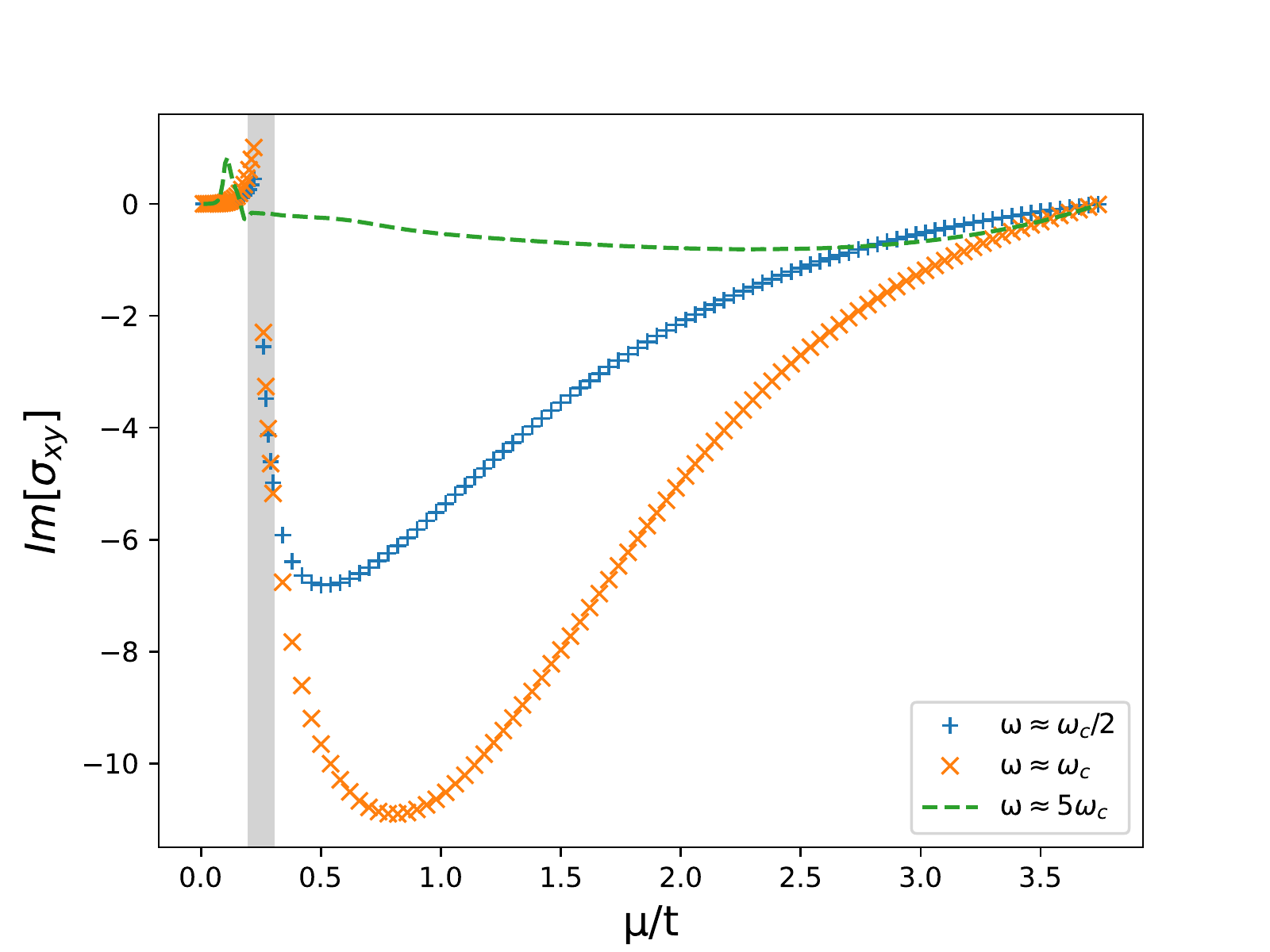}
\caption{Real and imaginary part of Hall conductivity for $\omega_{c}\tau_{S}=1$.}
\label{fig:Graphene_sigma_xy_tau=00003D1} 
\end{figure}

\begin{figure}[!]
\includegraphics[width=0.4\textwidth]{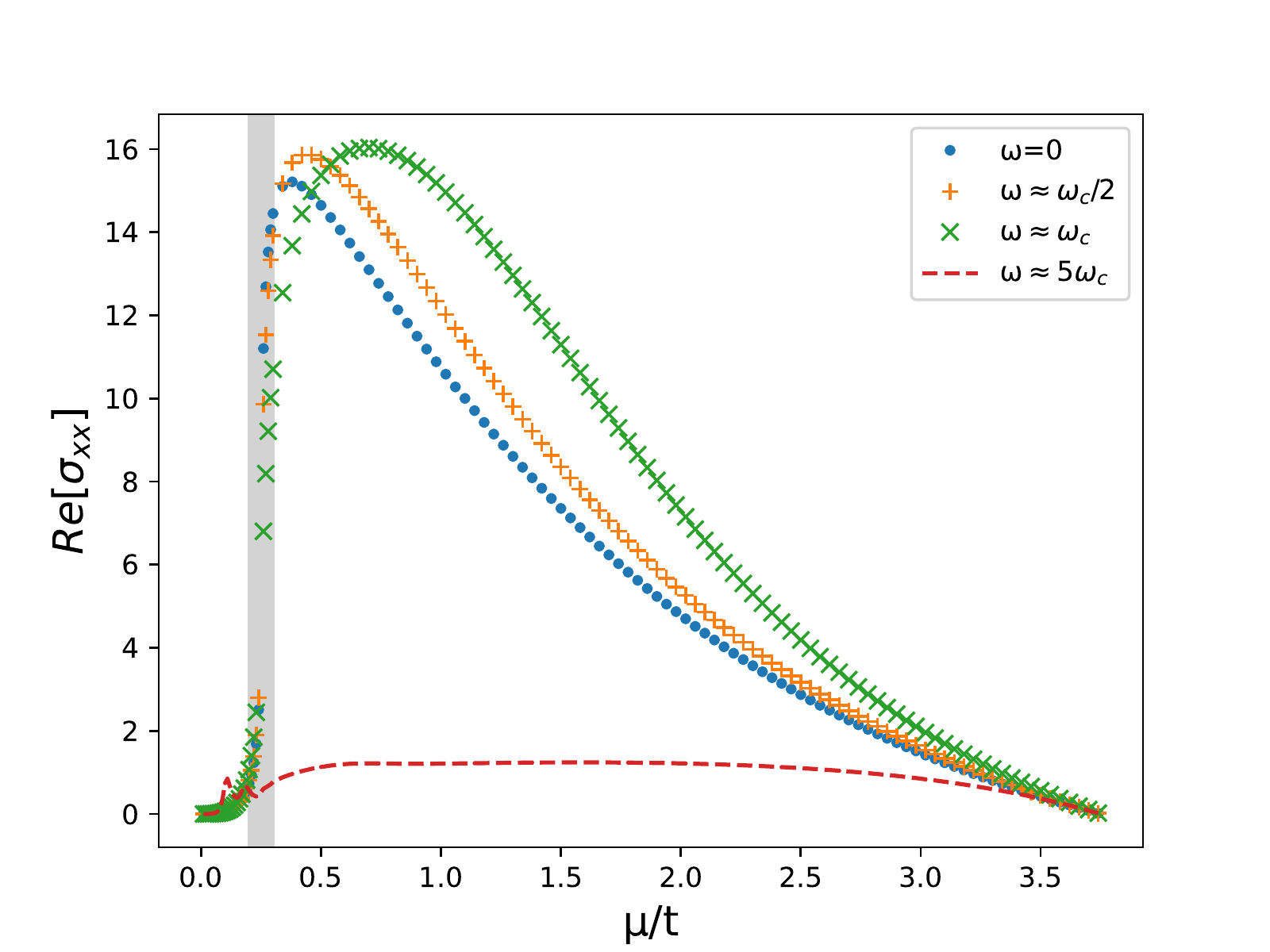}
\includegraphics[width=0.4\textwidth]{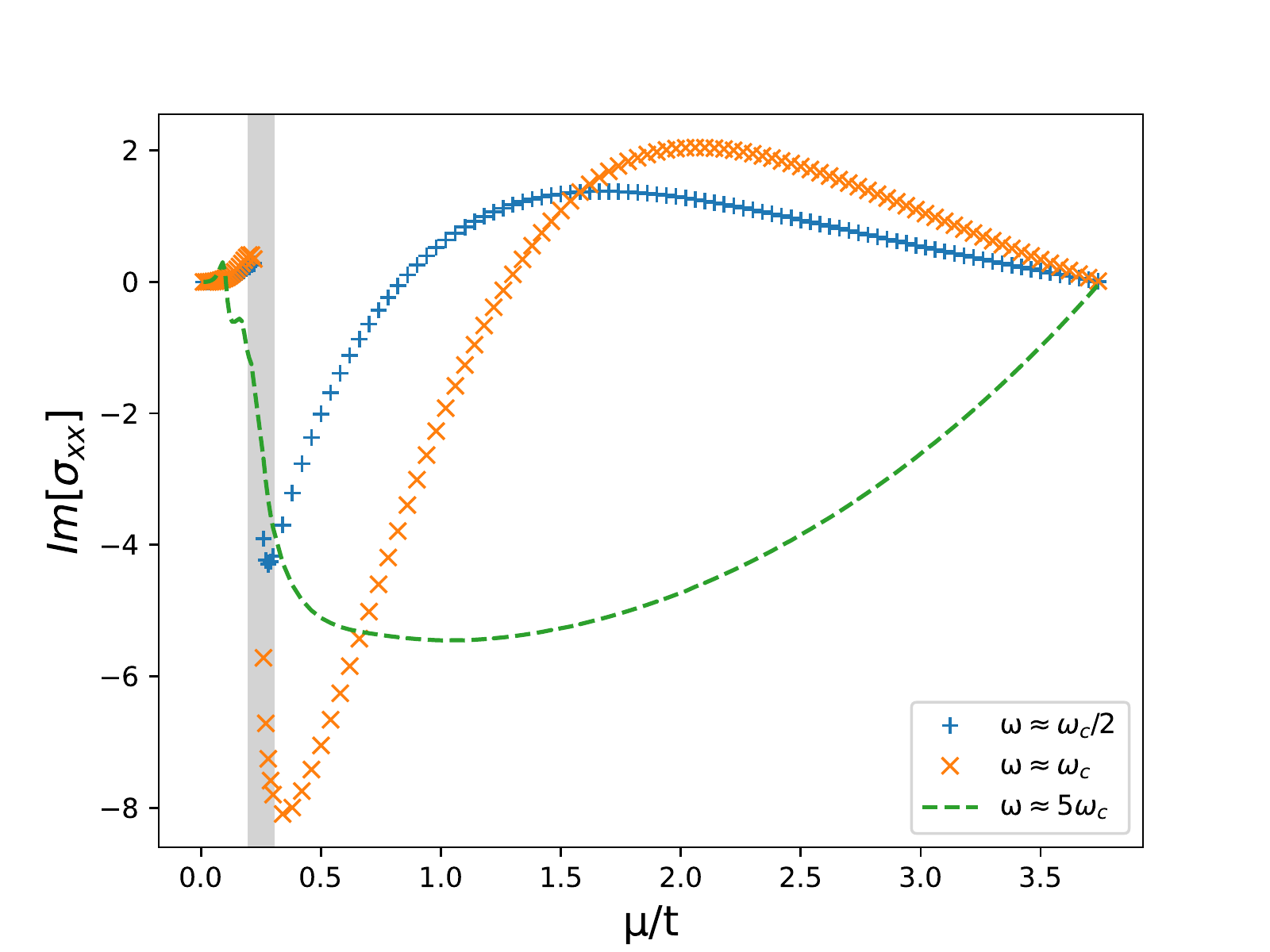}
 \caption{Real and imaginary part of the conductivity $\sigma_{xx}$ (same as
$\sigma_{yy}$) for $\omega_{c}\tau_{S}=1$.}
\label{fig:Graphene_sigma_xx_yy_tau=00003D1} 
\end{figure}

The numerical results for the Hall coefficient as well as the conductivity
components are presented in Figs (4-9). From the scaling formula Eq.
(\ref{eq:Scaling-3}) $T\tau_{S}\rightarrow0$ and we take $eB\rightarrow0.01$
for the calculations. In this case $\omega_{c}$ is defined as $\omega_{c}=\frac{eB}{m^{*c}}$
with the effective mass $m^{*}\propto t^{-1}$. 
We have chosen again two values of $\omega_{c}\tau_{S}$ (0.3 and
1) and for each one three frequencies $\omega$ for the time-dependent
electric field. For c/t=1/4, the higher order Van Hove singularity
is at $\mu/t=0.25$. This explains the sharp features of the conductivities
and the change of sign of the Hall coefficient at that value. In Fig.
(5a) the Fermi surface topology is depicted across both sides of the
discontinuous behavior. The maximum of the real part of the conductivity
takes place when the area of the Fermi surface is largest. The frequency
dependence is pronounced in the imaginary part of the Hall coefficient
$R_{H}$ and around the value of $\mu/t=0.25$. Away from $\mu/t=0.25$
the imaginary part of $R_{H}$ is zero, while for the real part of
$R_{H}$ the behavior is linear in $\mu/t$. This result is very well
explained by the fact that away from the FSTT,
the dispersion relation is very well approximated by a parabolic one
with precisely this behavior as the imaginary part of $\rho_{xy}$
is 0 and for the real part $\rho_{xy}\propto B/(ne)$, where $n$
is the density of electrons which is linear in $\mu$. As a result
$R_{H}$ exhibits that behavior for $\mu/t>1$. As the frequency is
increased the region where there is appreciable effect of the singularity
is larger.


\section{\label{sec:Hall-conductivity-of}Hall conductivity of the Haldane model}

\begin{figure}[!]
\centering
\includegraphics[width=0.4\textwidth]{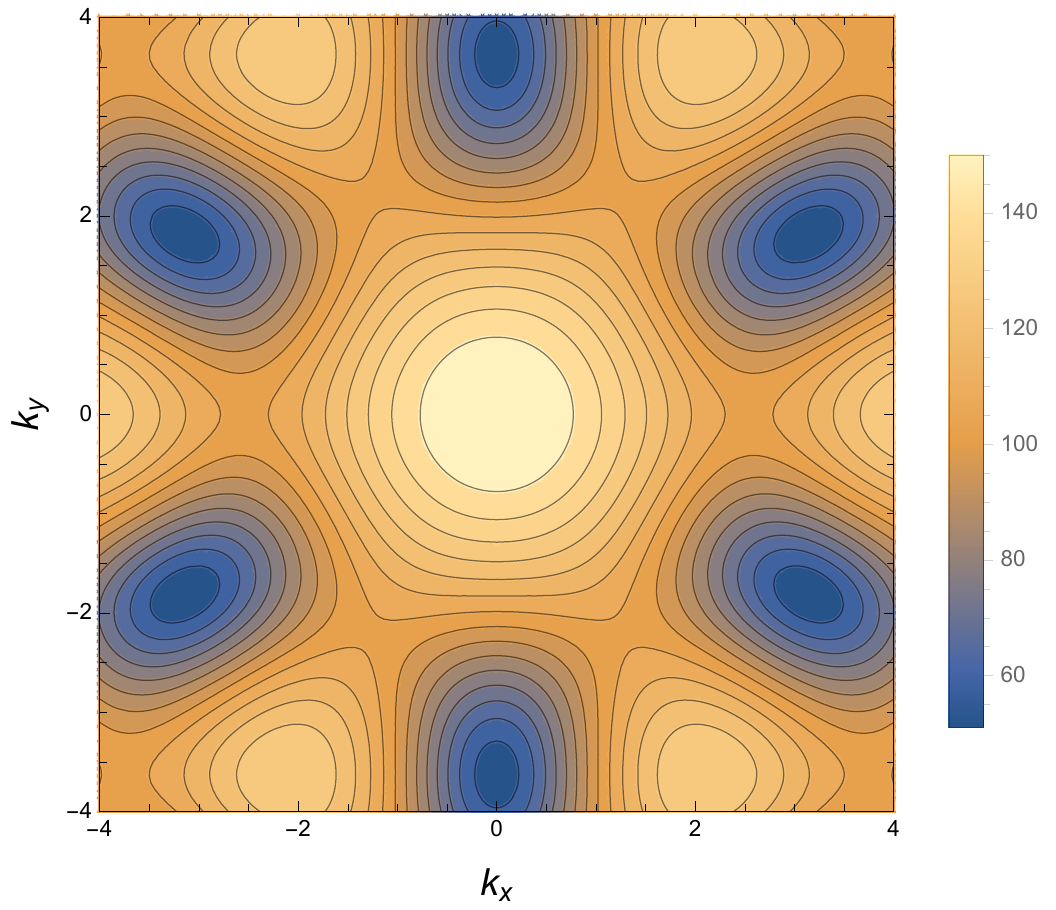}
\caption{Energy contours of the Haldane model, the parameters we use are mentioned
in the text.}
\label{fig:Haldane_Dispersion}
\end{figure}

We consider the Haldane Hamiltonian on a honeycomb lattice with $H(\theta)=d_{0}(\theta)+{\bf d}(\theta)\cdot{\bf \sigma}$
where: 

\begin{eqnarray*}
d_{0}(\theta) & = & -2t_{2}\left[\cos{\theta}_{1}+\cos{\theta}_{2}+\cos({\theta}_{1}+{\theta}_{2})\right]\cos\phi\\
d_{x}(\theta) & = & -t_{1}(1+\cos{\theta}_{1}+\cos{\theta}_{2})\\
d_{y}(\theta) & = & t_{1}(\sin{\theta}_{1}-\sin{\theta}_{2})\\
d_{z}(\theta) & = & m-2t_{2}\left[\sin{\theta}_{1}+\sin{\theta}_{2}-\sin({\theta}_{1}+{\theta}_{2})\right]\sin\phi
\end{eqnarray*}
and ${\bf \sigma}$ denotes the Pauli matrices. The energy eigenvalues
are given by $E_{\pm}=d_{0}\pm|d_{1}|$. For simplicity, we choose
$\phi=\pi/2$, $m=0$ and consider $E_{+}$ for $t_{1}=50$ and $t_{2}=25$
with a nontrivial Chern number of -1. The energy contours of the model
is presented in Fig. (\ref{fig:Haldane_Dispersion}). We numerically calculate the Hall conductivity
and present the results in Figs (11) - (14). The real part of the Hall
conductivity shows different behavior in different regimes of the
chemical potential reflecting the different Fermi surface topology.
This topology of the Fermi surface is shown in the insets of Fig.
(2), in each segment where the behavior is different. In the value
of the total Hall conductivity a constant $-\frac{1}{2\pi}$ has been
added in Fig. (2), due to the contribution of the lower band. The
results are for $\omega_{c}\tau_{S}=0.3$, where $\omega_{c}=\frac{eB}{m^{*}c}$
and the effective mass $m^{*}\propto{t_{1}}^{-1}$ and $eB=0.01$.
The contributions from the different terms of Eq.(\ref{eq:Current-2-1-2})
are presented separately and the significant role of the second term
of Eq.(\ref{eq:Current-2-1-2}) is clearly demonstrated.

\begin{figure}[!]
\centering
\includegraphics[width=0.4\textwidth]{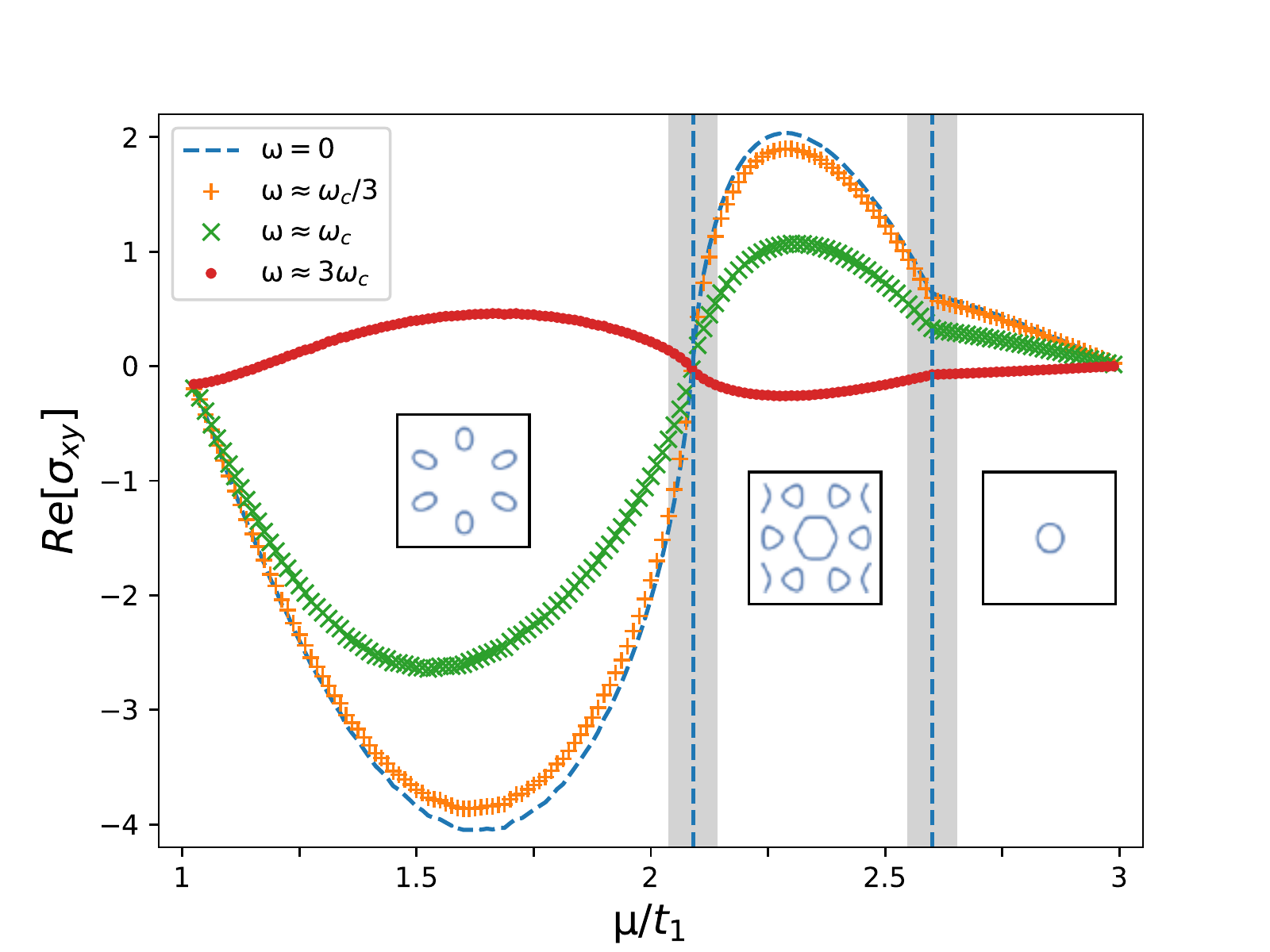}
\caption{The real part of the Hall conductivity. In the various parts of the
curve, the Fermi surface topology is shown. All conductivities are
in units of $e^{2}\tau_{S}$. For the second term of Eq.(18), which
we call "regular" term, we use the scaling relation:
$\sigma_{\alpha\beta}^{reg}\left(e,\tau_{S},T,\mathbf{B},\omega,\varepsilon_{M}\right)=e\tau_{S}^{-1}\sigma_{\alpha\beta}^{reg}\left(1,1,T\tau_{S},e\mathbf{B},\omega\tau_{S},\tau_{S}\varepsilon_{M}\right)$. As before, the shaded areas denote the places where magnetic breakdown phenomena should be taken into account for a full quantitative analysis.}
\includegraphics[width=0.4\textwidth]{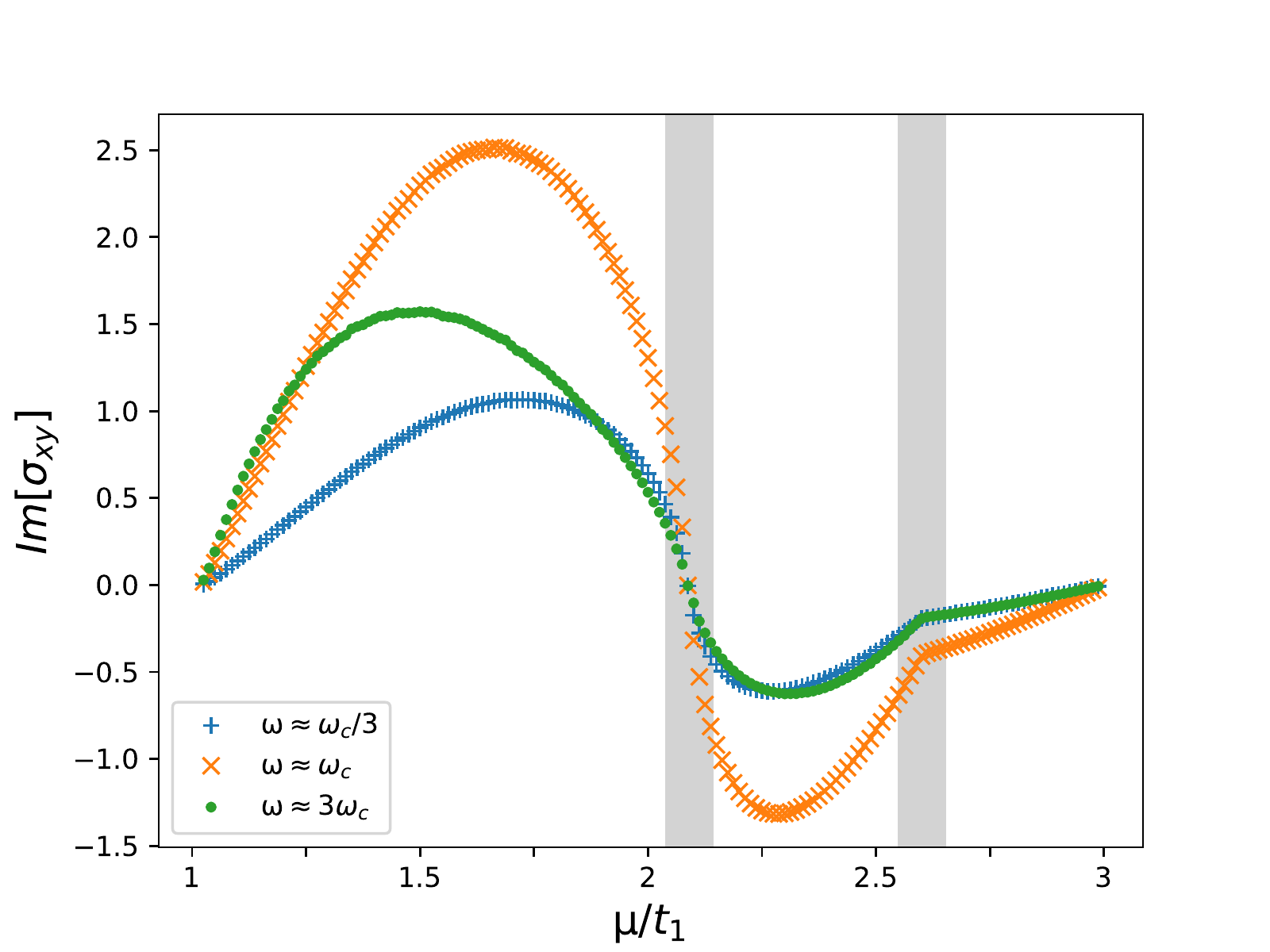}
\caption{Imaginary part of the Hall conductivity. This comes solely from the
second term of Eq. (18).}
\includegraphics[width=0.4\textwidth]{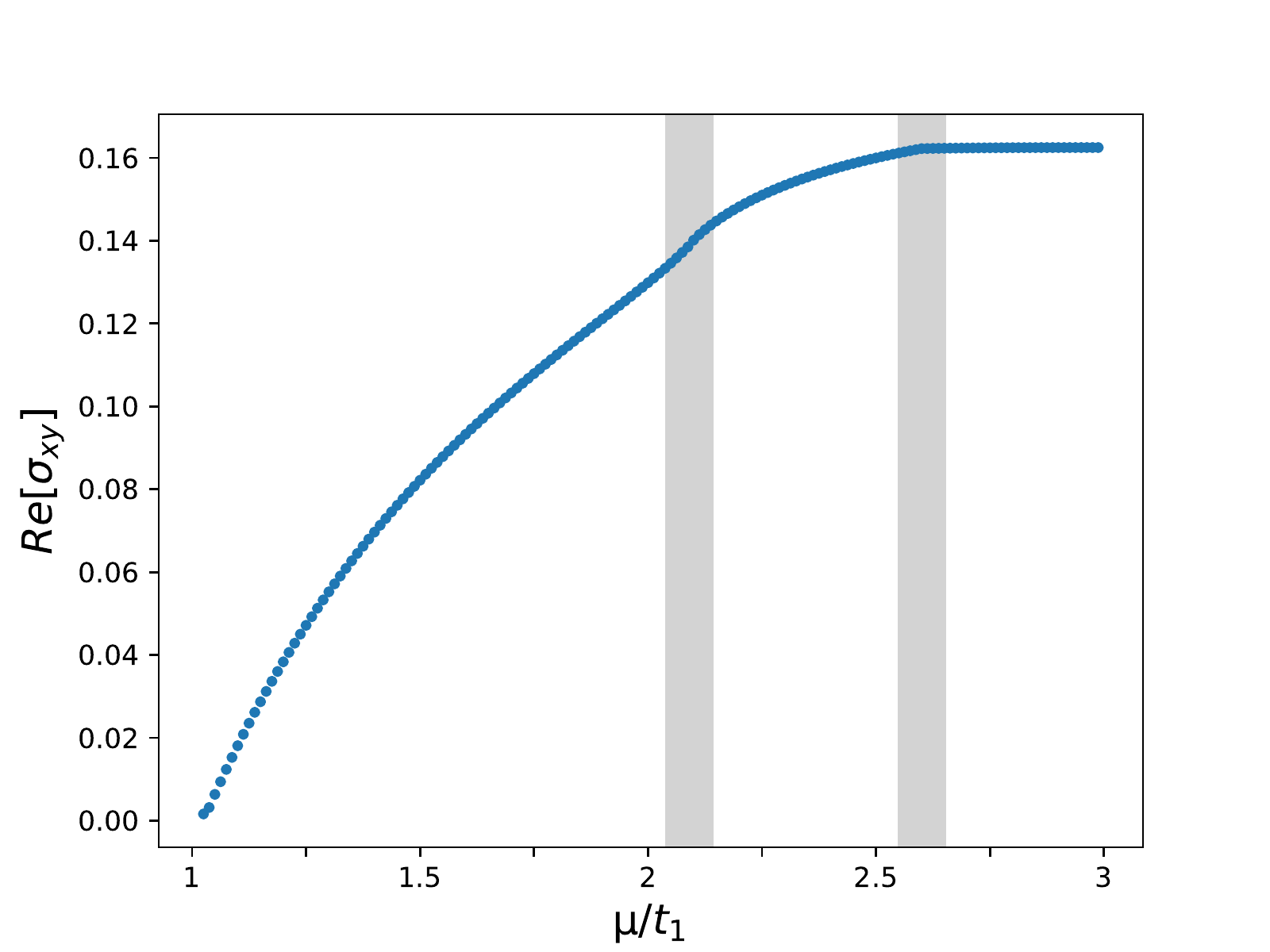} 
\caption{First term of Hall conductivity as given by Eq. (18). It has no frequency
dependence. The range of the values is from 0 to $\pi/2$, indicating
that, for $e=1$, the Chern number is -1. }
\end{figure}

\begin{figure}[!]
\centering
\includegraphics[width=0.4\textwidth]{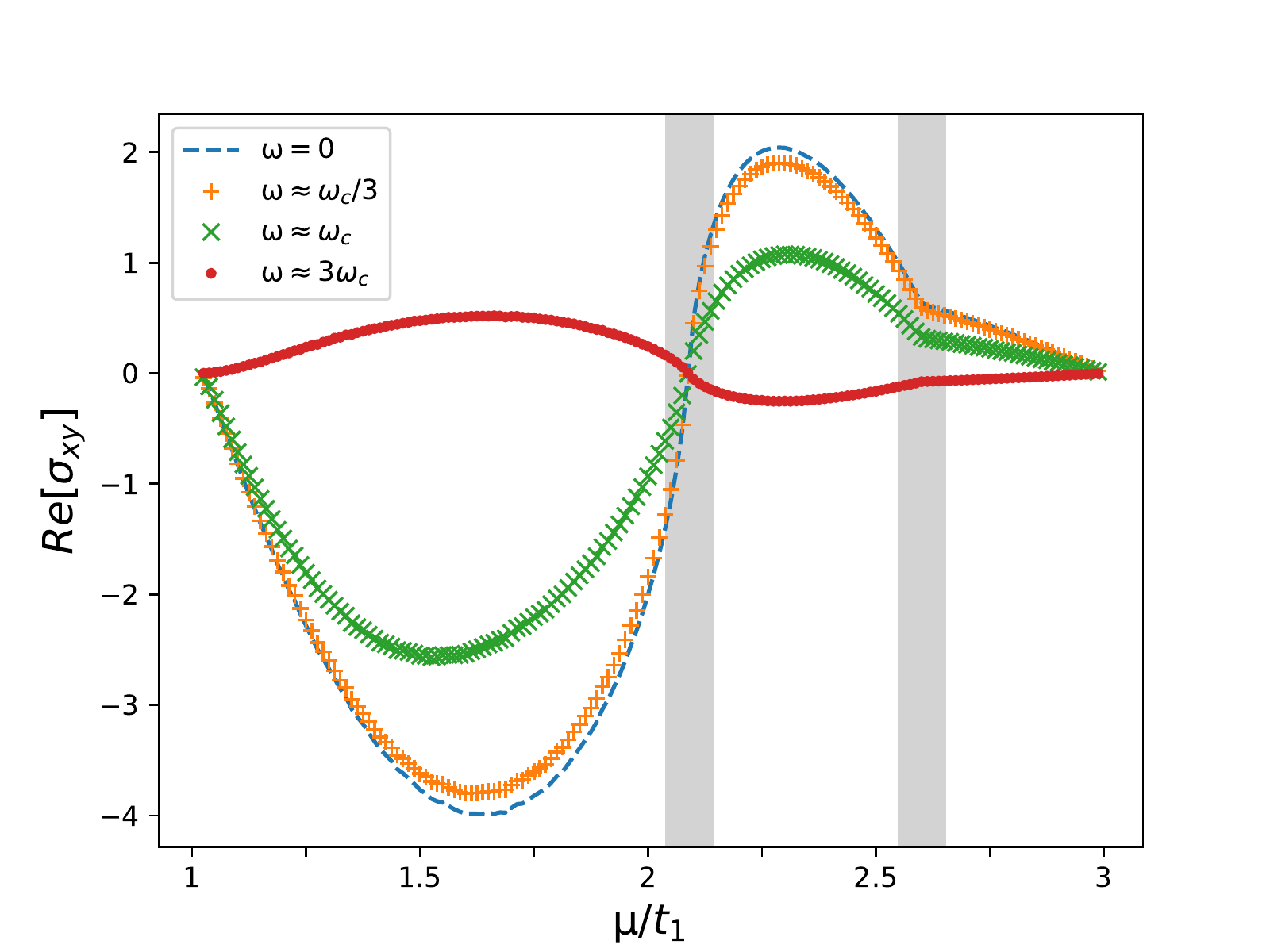}
\caption{The real part of the second term of Hall conductivity.}
\label{fig:Haldane_sigma_xy_2}
\end{figure}


\section{\label{sec:Conclusion}Conclusion}

In this work a generalization of the celebrated Chambers formula
has been introduced, relevant to time dependent electric fields and
bands with Berry curvature. The nonlinear conductivity, to order $E^{2}$
in the electric field has been also computed within the leading order
equations of motion method. These general formulae have been used
to study a number of examples. In particular, we studied bands where by changing the chemical potential a range of FSTTs become available.
These FSTTs lead to Van Hove
singularities at the Fermi surface where the high density of states
but our work is valid away from the regions close to the topological transitions to avoid considering quantum Hall or out-of-equilibrium effects.
Due to the change of the Fermi surface from hole like to electron
like there is a jump in the Hall coefficient of the material
at the Fermi surface topological transition, while a wealth of different
features and signatures appear both in real and imaginary parts of
the Hall coefficient and conductivities, especially pronounced at
higher frequencies. Different types of FSTTs provide their signatures on the conductivities.
Furthermore, the Hall conductivity of the Haldane model has been studied.

The main assumption of the general part of our work is that the system
is a Fermi liquid and the effects of the interactions can be included
in the lifetime and the effective mass. There is recent seminal work
that takes into account strong interactions \cite{Auerbach} and presents
a formula that does not require the existence of quasiparticles. We
leave as a future work how to bridge the two approaches. Given the
enormous interest in the field, we believe that this work will stimulate
significant future activity associated with applications of the extended,
formally exact Chambers formulae presented here. This work is based
on simplified semiclassical equations of motion that are commonly
used but are only strictly valid to leading order in the electric
and magnetic fields. As it has been already emphasized, close to Van Hove singularities, magnetic breakdown effects (quantum tunnelling), must be taken into account \cite{Falicov, Glazman}. These will be considered elsewhere.

\textit{Acknowledgments.} We would like to thank Claudio Chamon, Anirudh
Chandrasekaran, Mark Greenaway and Xenophon Zotos for useful discussions.
The work has been supported by the EPSRC grants EP/P002811/1 and EP/T034351/1
(JJB). DVE was supported by the RSF-DFG grant 405940956.

\appendix

\section{\label{sec:Usual-Chamberss-formula}Chambers formula}

\subsection{\label{sec:Boltzmann_Equation}Boltzmann Equation }

Following the Chambers original work  \cite{Chambers_1952} we write the current 
as:
\begin{equation}
J_{\alpha}\left(t\right)=-e\int\frac{d^{3}k}{\left(2\pi\right)^{3}}\frac{d\mathbf{r}_{\alpha}}{dt}\left({\bf k}\right)f\left({\bf k},t\right)\label{eq:Solution-4-1-1-1-1}
\end{equation}
with $f\left({\bf k},t\right)$ being the solution to the Boltzmann
equation: 
\begin{equation}
\frac{\partial f\left(\mathbf{k},t\right)}{\partial t}+\frac{d{\bf k}}{dt}\frac{\partial f\left(\mathbf{k},t\right)}{\partial{\bf k}}=\frac{1}{\tau_{S}\left(\mathbf{k}\right)}\left(f_{0}\left(\varepsilon_{M}\left(\mathbf{k}\right)\right)-f\left(\mathbf{k},t\right)\right)\label{eq:Boltzmann-1-2}
\end{equation}
with 
\begin{align}
\frac{d{\bf k}}{dt} & =-e\left[{\bf \nabla_{k}}\varepsilon\left({\bf k}\right)\times{\bf B}+{\bf E}\left(t\right)\right]\label{eq:Equation-2}\\
\frac{d{\bf r}}{dt} & =\nabla_{\mathbf{k}}\varepsilon\left({\bf k}\right)\label{eq:Velocity-1}
\end{align}
The solution as is well known, is given by: 
\begin{equation}
f\left({\bf k},t\right)=\int_{-\infty}^{t}\frac{f_{0}\left(\varepsilon\left({\bf k}(t')\right)\right)}{\tau_{S}\left(\mathbf{k}\left(t'\right)\right)}\exp\left(-\int_{t'}^{t}\frac{ds}{\tau_{S}\left(\mathbf{k}\left(s\right)\right)}\right)dt'\label{eq:Solution-5}
\end{equation}

\subsection{\label{subsec:Current-1}Momentum as a function of time}

\noindent In the limit where $\mathbf{E}\left(t\right)$ is small
the second term on the right hand side of Eq. (\ref{eq:Equation-2})
may be viewed as a perturbation. In that spirit let us denote by $\mathbf{k}_{0}\left(t\right)$
the solution to the equation: 
\begin{equation}
\frac{d\mathbf{k_{0}}\left(t\right)}{dt}=-e{\bf \nabla_{k}}\varepsilon\left({\bf k}_{0}\left(t\right)\right)\times{\bf B}\label{eq:Zero_olution-1-2}
\end{equation}
If we write: $\mathbf{E}\left(t\right)=\lambda\mathbf{E}\left(t\right)$
with $\lambda=1$, then the solution to Eq. (\ref{eq:Equation-2})
can be written as: 
\begin{equation}
\mathbf{k}\left(t\right)=\mathbf{k}_{0}\left(t\right)+\lambda\mathbf{k}_{1}\left(t\right)+\lambda^{2}\mathbf{k}_{2}\left(t\right)+....\label{eq:Solution-3-2}
\end{equation}
with $\mathbf{k}_{1}\left(t_{0}\right)=\mathbf{k}_{2}\left(t_{0}\right)=\mathbf{k}_{3}\left(t_{0}\right)=...=0$.
Then, to order $\lambda$ we obtain: 
\begin{equation}
\frac{d\mathbf{k}_{1}\left(t\right)}{dt}=-e\left[\sum_{\alpha}{k}_{1\alpha}\left(t\right)\frac{\partial}{\partial{k}_{\alpha}}{\bf \nabla_{\mathbf{k}}}\varepsilon\left({\bf k}_{0}\left(t\right)\right)\right]\times{\bf B}-e{\bf E}\left(t\right)\label{eq:K_1}
\end{equation}
In addition: 
\begin{eqnarray}
\frac{d\varepsilon\left(\mathbf{k}\left(t\right)\right)}{dt} & = & \nabla_{\mathbf{k}}\varepsilon\left(\mathbf{k}\left(t\right)\right)\cdot\frac{d\mathbf{k}}{dt}\nonumber \\
 & = & -e\nabla_{\mathbf{k}}\varepsilon\left(\mathbf{k}\right)\cdot\left[{\bf \nabla_{k}}\varepsilon\left({\bf k}\right)\times{\bf B}+{\bf E}\left(t\right)\right]\nonumber \\
 & = & -e\nabla_{\mathbf{k}}\varepsilon\left(\mathbf{k}\left(t\right)\right)\cdot{\bf E}\left(t\right)\label{eq:K_1_bar-1}
\end{eqnarray}

\noindent We can now similarly expand in powers of $\lambda$:

\begin{equation}
\varepsilon_{M}\left(t\right)=\varepsilon_{0}\left(t\right)+\lambda\varepsilon_{1}\left(t\right)+\lambda^{2}\varepsilon_{2}\left(t\right)+...\label{eq:Transform-2-1}
\end{equation}
with $\varepsilon_{1}\left(t_{0}\right)=\varepsilon_{2}\left(t_{0}\right)=\varepsilon_{3}\left(t_{0}\right)=...=0$.
To order $\lambda$ for $t>t_{0}$: 
\begin{equation}
\varepsilon_{1}\left(t\right)=e\int_{t_0}^{t}\nabla_{\mathbf{k}}\varepsilon\left(\mathbf{k}_{0}\left(s\right)\right)\cdot{\bf E}\left(s\right)ds\label{eq:Solution-1-2}
\end{equation}
and 
\begin{equation}
\varepsilon_{2}\left(t\right)=e\sum_{\alpha\beta}\int_{t_0}^{t}{E}_{\beta}\left(s\right){k}_{1\alpha}\left(s\right)\frac{\partial^{2}}{\partial{k}_{\alpha}\partial{k}_{\beta}}\varepsilon\left(\mathbf{k}_{0}\left(s\right)\right)ds\label{eq:E_2}
\end{equation}

\subsection{\label{sec:Main-calculation-2} Calculation of the current in linear
response in 3D}
We first note the identity: 
\begin{equation}
\int_{-\infty}^{t}\frac{dt'}{\tau_{S}\left(\mathbf{k}\left(t'\right)\right)}\exp\left(-\int_{t'}^{t}\frac{ds}{\tau_{S}\left(\mathbf{k}\left(s\right)\right)}\right)=1,\label{eq:Identity}
\end{equation}
therefore there is no need to expand the term $\int_{-\infty}^{t}\frac{dt'}{\tau_{S}\left(\mathbf{k}\left(t'\right)\right)}\exp\left(-\int_{t'}^{t}\frac{ds}{\tau_{S}\left(\mathbf{k}\left(s\right)\right)}\right)$
in Eq. (\ref{eq:Solution-5}). As a result we obtain: 
\begin{eqnarray}
\nonumber
f\left({\bf k},t\right)&=&f_{0}\left(\varepsilon\left(\mathbf{k}_{0}(t)\right)\right)\\
\nonumber
&+&\frac{\partial f_{0}\left(\varepsilon\left(\mathbf{k}_{0}(t)\right)\right)}{\partial\varepsilon}\int_{-\infty}^{t}dt'\frac{\varepsilon_{1}\left(t'\right)}{\tau_{S}\left(\mathbf{k}_{0}\left(t'\right)\right)}\eta(t;t')\label{eq:Density}
\end{eqnarray}
Noticing that the current is written as: 
\begin{equation}
J_{\alpha}\left(t\right)
%
=\frac{\partial f_{0}\left(\varepsilon\left(\mathbf{k}_{0}(t)\right)\right)}{\partial\varepsilon}\int_{-\infty}^{t}\frac{\varepsilon_{1}\left(t'\right)}{\tau_{S}\left(\mathbf{k}_{0}\left(t'\right)\right)}\eta(t;t')dt'
\label{eq:Current}
\end{equation}
to obtain the result to first order in the electric field (linear
response), all terms $\sim E^{2}$ are neglected and a term is dropped
due to the relation: 
$-e\int\frac{d^{3}k}{\left(2\pi\right)^{3}}\nabla_{\mathbf{k}}\varepsilon\left({\bf k}\right)f_{0}\left(\varepsilon\left(\mathbf{k}_{0}(t)\right)\right)=0$
as there is no current without an electric field. Then: 
\begin{widetext}
\begin{equation}
J_{\alpha}\left(t\right)=-e\int\frac{d^{3}k}{\left(2\pi\right)^{3}}\nabla_{{k}_{\alpha}}\varepsilon\left({\bf k}\right)\cdot\left[\frac{\partial f_{0}\left(\varepsilon\left(\mathbf{k}_{0}(t)\right)\right)}{\partial\varepsilon}\int_{-\infty}^{t}dt'\frac{\varepsilon_{1}\left(t'\right)}{\tau_{S}\left(\mathbf{k}_{0}\left(t'\right)\right)}\eta(t;t')\right]\label{eq:final_answer}
\end{equation}
The first term in the bracket in Eq. (\ref{eq:final_answer}) is what
Chambers calculated and it is given by \citep{Chambers_1952, Shockley_1950},
we can integrate it by parts using $\frac{d}{dt}\varepsilon_{1}\left(t\right)=-e\nabla_{\mathrm{\mathbf{k}}}\varepsilon\left(\mathbf{k}_{0}\left(t\right)\right)\cdot\mathbf{E}\left(t\right)$
to obtain the final expression: 
\begin{equation}
J_{\alpha}\left(t\right)=-e^{2}\int\frac{d^{3}k}{\left(2\pi\right)^{3}}\nabla_{{k}_{\alpha}}\varepsilon\left({\bf k}\right)\frac{\partial f_{0}\left(\varepsilon\left(\mathbf{k}\right)\right)}{\partial\varepsilon}\int_{-\infty}^{t}dt'\nabla_{\mathbf{k}}\varepsilon\left(\mathbf{k}_{0}(t')\right)\cdot\mathbf{E}\left(t\right)\eta(t;t')\label{eq:Current_1}
\end{equation}

\subsection{\label{sec:Order--terms-1}Calculation of nonlinear current to order
$\mathbf{E}^{2}$}

\noindent To proceed with the next order term, we note that: 
\begin{align}
f\left({\bf k},t\right) & \simeq\int_{-\infty}^{t}\frac{f_{0}\left(\left(\left[\varepsilon_{0}+\varepsilon_{1}+\varepsilon_{2}\right](t')\right)\right)}{\tau_{S}\left(\left[\mathbf{k}_{0}+\mathbf{k}_{1}+\mathbf{k}_{2}\right]\left(t'\right)\right)}\exp\left(-\int_{t'}^{t}\frac{ds}{\tau_{S}\left(\left[\mathbf{k}_{0}+\mathbf{k}_{1}+\mathbf{k}_{2}\right]\left(s\right)\right)}\right)dt'\nonumber \\
 & \cong\int_{-\infty}^{t}\frac{f_{0}\left(\varepsilon\left(\mathbf{k}_{0}(t)\right)\right)}{\tau_{S}\left(\mathbf{k}_{0}\left(t'\right)\right)}\eta(t;t')dt' +\frac{\partial f_{0}\left(\varepsilon\left(\mathbf{k}_{0}(t)\right)\right)}{\partial\varepsilon}\int_{-\infty}^{t}\frac{\varepsilon_{1}\left(t'\right)+\varepsilon_{2}\left(t'\right)}{\tau_{S}\left(\mathbf{k}_{0}\left(t'\right)\right)}\eta(t;t')dt'\nonumber \\
 & +\frac{1}{2}\frac{\partial^{2}f_{0}\left(\varepsilon\left(\mathbf{k}_{0}(t)\right)\right)}{\partial\varepsilon^{2}}\int_{-\infty}^{t}\frac{\varepsilon_{1}^{2}\left(t'\right)}{\tau_{S}\left(\mathbf{k}_{0}\left(t'\right)\right)}\eta(t;t')dt' -\frac{\partial f_{0}\left(\varepsilon\left(\mathbf{k}_{0}(t)\right)\right)}{\partial\varepsilon}\int_{-\infty}^{t}\frac{\varepsilon_{1}\left(t\right)}{\tau_{S}\left(\mathbf{k}_{0}\left(t'\right)\right)}\frac{\nabla_{\mathbf{k}}\tau_{S}\left(\mathbf{k}_{0}(t')\right)\cdot\mathbf{k}_{1}(t')}{\tau_{S}\left(\mathbf{k}_{0}\left(t\right)\right)}\eta(t;t')dt'\nonumber \\
 & +\frac{\partial f_{0}\left(\varepsilon\left(\mathbf{k}_{0}(t)\right)\right)}{\partial\varepsilon}\int_{-\infty}^{t}\frac{\varepsilon_{1}\left(t\right)}{\tau_{S}\left(\mathbf{k}_{0}\left(t\right)\right)}\eta(t;t')\int_{t'}^{t}\frac{\nabla_{\mathbf{k}}\tau_{S}\left(\mathbf{k}_{0}(l)\right)\mathbf{k}_{1}(l)}{\tau_{S}^{2}\left(\mathbf{k}_{0}\left(l\right)\right)}dldt'\label{eq:Expansion}
\end{align}
where Eq. (\ref{eq:Identity}) has been used. Integrating the above
expression by parts and using the formula for the current: 
\begin{align}
J_{\alpha}^{(2)}\left(t\right) & =-e^{2}\int\frac{d^{3}k}{\left(2\pi\right)^{3}}\nabla_{{k}_{\alpha}}\varepsilon\left({\bf k}\right)\frac{\partial f_{0}\left(\varepsilon\left(\mathbf{k}\right)\right)}{\partial\varepsilon}\int_{-\infty}^{t}dt'\nabla_{\mathbf{k}}\varepsilon\left(\mathbf{k}_{0}(t')\right)\cdot\mathbf{E}\left(t\right)\eta(t;t')\nonumber \\
 & -e\int\frac{d^{3}k}{\left(2\pi\right)^{3}}\nabla_{{k}_{\alpha}}\varepsilon\left({\bf k}\right)\frac{\partial f_{0}\left(\varepsilon\left(\mathbf{k}\right)\right)}{\partial\varepsilon}\int_{-\infty}^{t}dt'\frac{\varepsilon_{2}\left(t'\right)}{\tau_{S}\left(\mathbf{k}_{0}\left(t'\right)\right)}\eta(t;t')\nonumber \\
 & -\frac{1}{2}e\int\frac{d^{3}k}{\left(2\pi\right)^{3}}\nabla_{{k}_{\alpha}}\varepsilon\left({\bf k}\right)\frac{\partial^{2}f_{0}\left(\varepsilon\left(\mathbf{k}\right)\right)}{\partial\varepsilon^{2}}\int_{-\infty}^{t}dt'\frac{\varepsilon_{1}^{2}\left(t'\right)}{\tau_{S}\left(\mathbf{k}_{0}\left(t'\right)\right)}\eta(t;t')\nonumber \\
 & -e^{2}\int\frac{d^{3}k}{\left(2\pi\right)^{3}}\nabla_{{k}_{\alpha}}\varepsilon\left({\bf k}\right)\frac{\partial f_{0}\left(\varepsilon\left(\mathbf{k}\right)\right)}{\partial\varepsilon}\int_{-\infty}^{t}dt'\nabla_{\mathbf{k}}\varepsilon\left(\mathbf{k}_{0}\left(t'\right)\right)\cdot{\bf E}\left(t'\right)\eta(t;t')\int_{t'}^{t}\frac{\nabla_{\mathbf{k}}\tau_{S}\left(\mathbf{k}_{0}(l)\right)\cdot\mathbf{k}_{1}(l)}{\tau_{S}^{2}\left(\mathbf{k}_{0}\left(l\right)\right)}dl\label{eq:Current_1-1}
\end{align}
finally, after another integration by parts, we obtain the expression
of the non-linear current: 
\begin{align}
J_{\alpha}^{(2)}\left(t\right) & =-e^{2}\int\frac{d^{3}k}{\left(2\pi\right)^{3}}\nabla_{{k}_{\alpha}}\varepsilon\left({\bf k}\right)\frac{\partial f_{0}\left(\varepsilon\left(\mathbf{k}\right)\right)}{\partial\varepsilon}\int_{-\infty}^{t}dt'\nabla_{\mathbf{k}}\varepsilon\left(\mathbf{k}_{0}(t')\right)\cdot\mathbf{E}\left(t\right)\eta(t;t')\nonumber \\
 & -e^{2}\int\frac{d^{3}k}{\left(2\pi\right)^{3}}\nabla_{{k}_{\alpha}}\varepsilon\left({\bf k}\right)\frac{\partial f_{0}\left(\varepsilon\left(\mathbf{k}\right)\right)}{\partial\varepsilon}\int_{-\infty}^{t}dt'\sum_{\beta\gamma}{E}_{\beta}\left(t'\right){k}_{1\gamma}\left(t'\right)\frac{\partial^{2}}{\partial{k}_{\gamma}\partial{k}_{\beta}}\varepsilon\left(\mathbf{k}_{0}\left(t'\right)\right)\eta(t;t')\nonumber \\
 & -e^{3}\frac{\partial^{2}f_{0}\left(\varepsilon\left(\mathbf{k}_{0}(t)\right)\right)}{\partial\varepsilon^{2}}\int_{-\infty}^{t}dt'\nabla_{\mathbf{k}}\varepsilon\left(\mathbf{k}_{0}(t')\right)\cdot\mathbf{E}\left(t'\right)\eta(t;t')\int_{t'}^{t}\nabla_{\mathbf{k}}\varepsilon\left(\mathbf{k}_{0}\left(l\right)\right)\cdot{\bf E}\left(l\right)dl\nonumber \\
 & -e^{2}\int\frac{d^{3}k}{\left(2\pi\right)^{3}}\nabla_{{k}_{\alpha}}\varepsilon\left({\bf k}\right)\frac{\partial f_{0}\left(\varepsilon\left(\mathbf{k}\right)\right)}{\partial\varepsilon}\int_{-\infty}^{t}dt'\nabla_{\mathbf{k}}\varepsilon\left(\mathbf{k}_{0}\left(t'\right)\right)\cdot{\bf E}\left(t'\right)\eta(t;t')\int_{t'}^{t}\frac{\nabla_{\mathbf{k}}\tau_{S}\left(\mathbf{k}_{0}(l)\right)\cdot\mathbf{k}_{1}(l)}{\tau_{S}^{2}\left(\mathbf{k}_{0}\left(l\right)\right)}dl\label{eq:Current_1-1-1}
\end{align}

\section{\label{sec:Main-calculation}Inclusion of Berry curvature}

\subsection{\label{sec:Order--terms-1} Calculation of the current in 3D in linear
response}

The distribution function is written as: 
\begin{equation}
f\left({\bf k},t\right)\simeq\int_{-\infty}^{t}\frac{dt}{\tau_{S}\left({\bf k}_{0}\left(t\right)\right)}f_{0}\left(\varepsilon_{M}\left(\mathbf{k}\left(t\right)\right)\right)\eta(t;t')+\frac{\partial f_{0}\left(\varepsilon_{M}\left(\mathbf{k}_{0}(t)\right)\right)}{\partial\varepsilon}\int_{-\infty}^{t}dt'\frac{\varepsilon_{1}\left(t'\right)}{\tau_{S}\left(\mathbf{k}_{0}\left(t'\right)\right)}\eta(t;t')\label{eq:Density-1}
\end{equation}
Substituting Eq. (\ref{eq:Density-1}) into Eq. (\ref{eq:Solution-4})
we get that the current is: 
\begin{align*}
J_{\alpha}\left(t\right) & =-e\int\frac{d^{3}k}{\left(2\pi\right)^{3}}\frac{dr_{\alpha}}{dt}D\left(\mathbf{k}\right)f\left(\mathbf{k},t\right)\\
 & =-e\int\frac{d^{3}k}{\left(2\pi\right)^{3}}D\left(\mathbf{k}\right)\left[u_{\mathbf{0}\alpha}\left(\mathbf{k}\right)+D^{-1}\left(\mathbf{k}\right)[e\mathbf{E}\left(t\right)\times{\bf \Omega}\left(k\right)]_{\alpha}\right]\times\\
 & \left[f_{0}\left(\varepsilon_{M}\left(\mathbf{k}\left(t\right)\right)\right)\int_{-\infty}^{t}\frac{dt'}{\tau_{S}\left({\bf k}_{0}\left(t'\right)\right)}\eta(t;t')+\frac{\partial f_{0}\left(\varepsilon_{M}\left(\mathbf{k}_{0}(t)\right)\right)}{\partial\varepsilon}\int_{-\infty}^{t}dt'\frac{\varepsilon_{1}\left(t'\right)}{\tau_{S}\left(\mathbf{k}_{0}\left(t'\right)\right)}\eta(t;t')\right]
\end{align*}
Following similar steps as before to obtain the linear response, we
get Eq. (\ref{eq:Current-2-1-1}) of the main text. 

\subsection{\label{subsec:Current-to-order}Calculation of nonlinear current
to order $E^{2}$ in 3D}

In this case, the distribution function reads: 
\begin{align}
f\left({\bf k},t\right) & \cong\int_{-\infty}^{t}\frac{f_{0}\left(\varepsilon\left(\mathbf{k}_{0}(t)\right)\right)}{\tau_{S}\left(\mathbf{k}_{0}\left(t'\right)\right)}\eta(t;t')dt' +\frac{\partial f_{0}\left(\varepsilon\left(\mathbf{k}_{0}(t)\right)\right)}{\partial\varepsilon}\int_{-\infty}^{t}\frac{\varepsilon_{1}\left(t'\right)+\varepsilon_{2}\left(t'\right)}{\tau_{S}\left(\mathbf{k}_{0}\left(t'\right)\right)}\eta(t;t')dt'\nonumber \\
 & +\frac{1}{2}\frac{\partial^{2}f_{0}\left(\varepsilon\left(\mathbf{k}_{0}(t)\right)\right)}{\partial\varepsilon^{2}}\int_{-\infty}^{t}\frac{\varepsilon_{1}^{2}\left(t'\right)}{\tau_{S}\left(\mathbf{k}_{0}\left(t'\right)\right)}\eta(t;t')dt'\nonumber \\
 & -\frac{\partial f_{0}\left(\varepsilon\left(\mathbf{k}_{0}(t)\right)\right)}{\partial\varepsilon}\int_{-\infty}^{t}dt'\frac{d\varepsilon_{1}\left(t'\right)}{dt}\eta(t;t')\int_{t'}^{t}\frac{\nabla_{\mathbf{k}}\tau_{S}\left(\mathbf{k}_{0}(l)\right)\cdot\mathbf{k}_{1}(l)}{\tau_{S}^{2}\left(\mathbf{k}_{0}\left(l\right)\right)}dl\label{eq:Distribution-1}
\end{align}
Then, we obtain for the current: 
\begin{align*}
J_{\alpha}&\left(t\right)  =-e\int\frac{d^{3}k}{\left(2\pi\right)^{3}}D\left(\mathbf{k}\right)\left[u_{\mathbf{0}\alpha}\left(\mathbf{k}\right)+D^{-1}\left(\mathbf{k}\right)e\left[\mathbf{E}\left(t\right)\times{\bf \Omega}\left(\mathbf{k}\right)\right]_{\alpha}\right]\times\\
 & \left[\int_{-\infty}^{t}dt'\frac{f_{0}\left(\varepsilon\left(\mathbf{k}_{0}(t)\right)\right)}{\tau_{S}\left(\mathbf{k}_{0}\left(t'\right)\right)}\exp\left(-\int_{t'}^{t}\frac{ds}{\tau_{S}\left(\mathbf{k}_{0}\left(s\right)\right)}\right) +\frac{\partial f_{0}\left(\varepsilon\left(\mathbf{k}_{0}(t)\right)\right)}{\partial\varepsilon}\int_{-\infty}^{t}dt'\frac{\varepsilon_{1}\left(t'\right)+\varepsilon_{2}\left(t'\right)}{\tau_{S}\left(\mathbf{k}_{0}\left(t'\right)\right)}\eta(t;t')\right.\\
 & +\left.\frac{1}{2}\frac{\partial^{2}f_{0}\left(\varepsilon\left(\mathbf{k}_{0}(t)\right)\right)}{\partial\varepsilon^{2}}\int_{-\infty}^{t}\frac{\varepsilon_{1}^{2}\left(t\right)}{\tau_{S}\left(\mathbf{k}_{0}\left(t\right)\right)}\eta(t;t')dt'-\frac{\partial f_{0}\left(\varepsilon\left(\mathbf{k}_{0}(t)\right)\right)}{\partial\varepsilon}\int_{-\infty}^{t}dt'\frac{d\varepsilon_{1}\left(t'\right)}{dt}\eta(t;t')\int_{t'}^{t}\frac{\nabla_{\mathbf{k}}\tau_{S}\left(\mathbf{k}_{0}(l)\right)\cdot\mathbf{k}_{1}(l)}{\tau_{S}^{2}\left(\mathbf{k}_{0}\left(l\right)\right)}dl\right]
\end{align*}
Focusing on terms $\propto E^{2}$ and using partial integration and
some simplifications, finally Eq.(\ref{eq:Current_E^2-3-2}) of the
main text is obtained. 

\subsection{\label{subsec:Main-calculation-1} Calculation of the current in 2D}

Having introduced ${\bf u_{0}^{2D}}$ in the main text and following
an identical derivation to Section \ref{subsec:Current-to-order}
the results in 2D for the nonlinear response reads: 
\begin{align}
 & J_{\alpha}^{\left(2\right)}\left(t\right)=\nonumber \\
 & -e^{2}\int\frac{d^{2}k}{\left(2\pi\right)^{2}}D\left(\mathbf{k}\right)\mathbf{u}_{\mathbf{0}\alpha}^{\mathbf{2D}}\left(\mathbf{k}\right)\left[\frac{\partial f_{0}\left(\varepsilon\left(\mathbf{k}_{0}(t)\right)\right)}{\partial\varepsilon}\int_{-\infty}^{t}dt'\sum_{\beta\gamma}{E}_{\beta}\left(t\right){k}_{1\gamma}\left(t'\right)\frac{\partial}{\partial{k}_{\gamma}}\left[D^{-1}\left(\mathbf{k}_{0}\left(t'\right)\right)\frac{\partial}{\partial{k}_{\beta}}\varepsilon\left(\mathbf{k}_{0}\left(t'\right)\right)\right]\eta(t;t')\right.\nonumber \\
 & +e\frac{\partial^{2}f_{0}\left(\varepsilon\left(\mathbf{k}_{0}(t)\right)\right)}{\partial\varepsilon^{2}}\int_{-\infty}^{t}dt'D^{-1}\left(\mathbf{k}_{0}\left(t'\right)\right)\nabla_{\mathbf{k}}\varepsilon\left(\mathbf{k}_{0}(t)\right)\cdot\mathbf{E}\left(t'\right)\eta(t;t')\int_{t'}^{t}D^{-1}\left(\mathbf{k}_{0}\left(l\right)\right)\nabla_{\mathbf{k}}\varepsilon\left(\mathbf{k}_{0}\left(l\right)\right)\cdot{\bf E}\left(l\right)dl\nonumber \\
 & \left.+\frac{\partial f_{0}\left(\varepsilon\left(\mathbf{k}_{0}(t)\right)\right)}{\partial\varepsilon}\int_{-\infty}^{t}dt'D^{-1}\left(\mathbf{k}_{0}\left(t'\right)\right)\nabla_{\mathbf{k}}\varepsilon\left(\mathbf{k}_{0}\left(t'\right)\right)\cdot{\bf E}\left(t'\right)\eta(t;t')\int_{t'}^{t}\frac{\nabla_{\mathbf{k}}\tau_{S}\left(\mathbf{k}_{0}(l)\right)\cdot\mathbf{k}_{1}(l)}{\tau_{S}^{2}\left(\mathbf{k}_{0}\left(l\right)\right)}dl\right]\nonumber \\
 & -e^{3}\varepsilon_{\alpha\beta}\int\frac{d^{2}k}{\left(2\pi\right)^{2}}{E}_{\beta}\left(t\right)\Omega\left(\mathbf{k}\right)\cdot\frac{\partial f_{0}\left(\varepsilon\left(\mathbf{k}\right)\right)}{\partial\varepsilon}\int_{-\infty}^{t}dt'D^{-1}\left(\mathbf{k}_{0}\left(t'\right)\right)\nabla_{\mathbf{k}}\varepsilon\left(\mathbf{k}_{0}(t')\right)\cdot\mathbf{E}\left(t'\right)\eta(t;t')\label{eq:Current_E^2-3-1}
\end{align}

\section{\label{sec:Hall-coeffcient-calculations}Hall coefficient calculations
for rectangular lattice}

In Ref. \citep{Maharaj_2017} the authors have solved Boltzmann transport
equations for this energy dispersion analytically. Then they calculated
the velocities $u_{x}$ and $u_{y}$ and took their Fourier series
expansions which are shown below:

\begin{equation}
u_{0x}^{i}\left(t\right)=\left(1-2\delta_{i,h}\right)\frac{2\pi}{m_{0}K\left(\kappa\right)}\sum_{n=1}^{\infty} sech\left[\frac{\left(2n-1\right)\pi K^{'}}{2K(\kappa)}\right]\sin\left[\frac{\left(2n-1\right)\pi u_{i}}{2K(\kappa)}\right]\sin\left[\frac{\left(2n-1\right)\pi\omega_{0}t}{2K\left(\kappa\right)}\right]
\end{equation}

\begin{equation}
u_{0y}^{i}\left(t\right)=\frac{2\pi}{m_{0}K\left(\kappa\right)}\sum_{n=1}^{\infty}sech\left[\frac{\left(2n-1\right)\pi K^{'}}{2K(\kappa)}\right]\cos\left[\frac{\left(2n-1\right)\pi u_{i}}{2K(\kappa)}\right]\cos\left[\frac{\left(2n-1\right)\pi\omega_{0}t}{2K\left(\kappa\right)}\right]
\end{equation}
where $i=e$ for electrons and $i=h$ for holes. For open surfaces
the corresponding results are:

\begin{equation}
u_{0x}^{o}\left(t\right)=\frac{2\pi\kappa}{m_{0}K\left(1/\kappa\right)}\sum_{n=1}^{\infty} sech\left[\frac{n\pi K^{'}}{K(1/\kappa)}\right]\sin\left[\frac{n\pi u_{o}}{K(1/\kappa)}\right]\sin\left[\frac{n\pi\kappa\omega_{0}t}{K\left(1/\kappa\right)}\right]
\end{equation}

\begin{equation}
u_{0y}^{o}\left(t\right)=\frac{2\pi\kappa}{m_{0}K\left(1/\kappa\right)}\left\{ \frac{1}{2}+\sum_{n=1}^{\infty}sech\left[\frac{n\pi K^{'}}{K(1/\kappa)}\right]\cos\left[\frac{n\pi u_{o}}{K(1/\kappa)}\right]\cos\left[\frac{n\pi\kappa\omega_{0}t}{K\left(1/\kappa\right)}\right]\right\} 
\end{equation}
where the definitions of $m_{0}$, $\kappa$, $\omega_{0}$, $K$
and $K^{'}$ are given in the main text.

For electron and hole pockets we can write for simplicity:

\begin{equation}
u_{0x}\left(t\right)=\tilde{u}_{x}\sum_{n=1}^{\infty}a_{n}^{x}\sin\left[\frac{\left(2n-1\right)\pi\omega_{0}t}{2K\left(\kappa\right)}\right]
\end{equation}

\begin{equation}
u_{0y}\left(t\right)=\tilde{u}_{y}\sum_{n=1}^{\infty}a_{n}^{y}\cos\left[\frac{\left(2n-1\right)\pi\omega_{0}t}{2K\left(\kappa\right)}\right]
\end{equation}
Eq. (17) is also true for open surfaces. But in the case of $u_{y}\left(t\right)$
we should write

\begin{equation}
u_{x}^{o}\left(t\right)=\tilde{u}_{x}\sum_{n=1}^{\infty}a_{n}^{x}\sin\left[\frac{n\pi\kappa\omega_{0}t}{K\left(1/\kappa\right)}\right]
\end{equation}

\begin{equation}
u_{y}^{o}\left(t\right)=\tilde{u}_{y}\left\{ \frac{1}{2}+\sum_{n=1}^{\infty}a_{n}^{y}\cos\left[\frac{n\pi\kappa\omega_{0}t}{K\left(1/\kappa\right)}\right]\right\} 
\end{equation}

Both for holes and electrons the results read for closed Fermi surfaces:

\begin{align}
\sigma_{xy} & =\frac{e^{3}B}{\left(2\pi\right)^{2}}\intop_{0}^{4K/\omega_{0}}\tilde{u}_{x}\sum_{n=1}^{\infty}a_{n}^{x}\sin\left[\frac{\left(2n-1\right)\pi\omega_{0}t}{2K\left(\kappa\right)}\right]dt\intop_{-\infty}^{t}\tilde{u}_{y}\sum_{m=1}^{\infty}a_{m}^{y}\cos\left[\frac{\left(2m-1\right)\pi\omega_{0}t'}{2K\left(\kappa\right)}\right]\exp\left(-\left[1/\tau_{S}+i\omega\right]\left(t-t'\right)\right)dt'\nonumber \\
 & =\frac{e^{3}B}{4\pi}\tilde{u}_{x}\tilde{u}_{y}\sum_{n=1}^{\infty}a_{n}^{x}a_{n}^{y}\frac{2n-1}{\left[1/\tau_{S}+i\omega\right]^{2}+\left[\frac{\left(2n-1\right)\pi\omega_{0}}{2K\left(\kappa\right)}\right]^{2}}\label{eq:Sigma_xy}
\end{align}

\begin{align}
\sigma_{xx} & =\frac{e^{3}B}{\left(2\pi\right)^{2}}\intop_{0}^{4K/\omega_{0}}\tilde{u}_{x}\sum_{n=1}^{\infty}a_{n}^{x}\sin\left[\frac{\left(2n-1\right)\pi\omega_{0}t}{2K\left(\kappa\right)}\right]dt\intop_{-\infty}^{t}\tilde{u}_{x}\sum_{m=1}^{\infty}a_{m}^{x}\sin\left[\frac{\left(2m-1\right)\pi\omega_{0}t'}{2K\left(\kappa\right)}\right]\exp\left(-\left[1/\tau_{S}+i\omega\right]\left(t-t'\right)\right)dt'\nonumber \\
 & =\frac{e^{3}B}{\left(2\pi\right)^{2}}\tilde{u}_{x}^{2}\sum_{n=1}^{\infty}\left(a_{n}^{x}\right)^{2}\frac{\left[1/\tau_{S}+i\omega\right]}{\left[1/\tau_{S}+i\omega\right]^{2}+\left[\frac{\left(2n-1\right)\pi\omega_{0}}{2K\left(\kappa\right)}\right]^{2}}\frac{2K\left(\kappa\right)}{\omega_{0}}\label{eq:Sigma_xx}
\end{align}

\begin{align}
\sigma_{yy} & =\frac{e^{3}B}{\left(2\pi\right)^{2}}\intop_{0}^{4K/\omega_{0}}\tilde{u}_{y}\sum_{n=1}^{\infty}a_{n}^{y}\cos\left[\frac{\left(2n-1\right)\pi\omega_{0}t}{2K\left(\kappa\right)}\right]dt\intop_{-\infty}^{t}\tilde{u}_{y}\sum_{m=1}^{\infty}a_{m}^{y}\cos\left[\frac{\left(2m-1\right)\pi\omega_{0}t'}{2K\left(\kappa\right)}\right]\exp\left(-\left[1/\tau_{S}+i\omega\right]\left(t-t'\right)\right)dt'\nonumber \\
 & =\frac{e^{3}B}{\left(2\pi\right)^{2}}\tilde{u}_{y}^{2}\sum_{n=1}^{\infty}\left(a_{n}^{y}\right)^{2}\frac{\left[1/\tau_{S}+i\omega\right]}{\left[1/\tau_{S}+i\omega\right]^{2}+\left[\frac{\left(2n-1\right)\pi\omega_{0}}{2K\left(\kappa\right)}\right]^{2}}\frac{2K\left(\kappa\right)}{\omega_{0}}\label{eq:Sigma_yy-1}
\end{align}

For open Fermi surfaces:

\begin{align}
\sigma_{xx} & =\frac{e^{3}B}{\left(2\pi\right)^{2}}\intop_{0}^{4K/\kappa\omega_{0}}\tilde{u}_{x}\sum_{n=1}^{\infty}a_{n}^{x}\sin\left[\frac{n\pi\kappa\omega_{0}t}{K\left(1/\kappa\right)}\right]dt\intop_{-\infty}^{t}\tilde{u}_{x}\sum_{m=1}^{\infty}a_{m}^{x}\sin\left[\frac{m\pi\kappa\omega_{0}t}{K\left(1/\kappa\right)}\right]\exp\left(-\left[1/\tau_{S}+i\omega\right]\left(t-t'\right)\right)dt'\nonumber \\
 & =\frac{e^{3}B}{\left(2\pi\right)^{2}}\tilde{u}_{x}^{2}\sum_{n=1}^{\infty}\left(a_{n}^{x}\right)^{2}\frac{\left[1/\tau_{S}+i\omega\right]}{\left[1/\tau_{S}+i\omega\right]^{2}+\left[\frac{n\pi\kappa\omega_{0}}{K\left(1/\kappa\right)}\right]^{2}}\frac{K\left(1/\kappa\right)}{\kappa\omega_{0}}\label{eq:sigma_xx-1}
\end{align}

\begin{align}
\sigma_{xy} & =\frac{e^{3}B}{\left(2\pi\right)^{2}}\intop_{0}^{4K/\kappa\omega_{0}}\tilde{u}_{x}\sum_{n=1}^{\infty}a_{n}^{x}\sin\left[\frac{n\pi\kappa\omega_{0}t}{K\left(1/\kappa\right)}\right]dt\intop_{-\infty}^{t}\tilde{u}_{y}\left\{ \frac{1}{2}+\sum_{m=1}^{\infty}a_{m}^{y}\cos\left[\frac{m\pi\kappa\omega_{0}t}{K\left(1/\kappa\right)}\right]\right\} \exp\left(-\left[1/\tau_{S}+i\omega\right]\left(t-t'\right)\right)dt'\nonumber \\
 & =\frac{e^{3}B}{\left(2\pi\right)^{2}}\tilde{u}_{x}\tilde{u}_{y}\sum_{n=1}^{\infty}a_{n}^{x}a_{n}^{y}\frac{n\pi}{\left[1/\tau_{S}+i\omega\right]^{2}+\left[\frac{n\pi\kappa\omega_{0}}{K\left(1/\kappa\right)}\right]^{2}}\label{eq:sigma_xy-1}
\end{align}

\begin{align}
\sigma_{yy} & =\frac{e^{3}B}{\left(2\pi\right)^{2}}\intop_{0}^{4K/\kappa\omega_{0}}\tilde{u}_{y}\left\{ \frac{1}{2}+\sum_{n=1}^{\infty}a_{n}^{y}\cos\left[\frac{n\pi\kappa\omega_{0}t}{K\left(1/\kappa\right)}\right]\right\} dt\intop_{-\infty}^{t}\tilde{u}_{y}\left\{ \frac{1}{2}+\sum_{m=1}^{\infty}a_{m}^{y}\cos\left[\frac{m\pi\kappa\omega_{0}t}{K\left(1/\kappa\right)}\right]\right\} \exp\left(-\left[1/\tau_{S}+i\omega\right]\left(t-t'\right)\right)dt'\nonumber \\
 & =\frac{e^{3}B}{\left(2\pi\right)^{2}}\frac{\tilde{u}_{y}^{2}}{\left[1/\tau_{S}+i\omega\right]}\frac{K\left(1/\kappa\right)}{\kappa\omega_{0}}+\frac{e^{3}B}{\left(2\pi\right)^{2}}\tilde{u}_{y}^{2}\sum_{n,1}^{\infty}\left(a_{n}^{y}\right)^{2}\frac{\left[1/\tau_{S}+i\omega\right]}{\left[1/\tau_{S}+i\omega\right]^{2}+\left[\frac{n\pi\kappa\omega_{0}}{K\left(1/\kappa\right)}\right]^{2}}\frac{K\left(1/\kappa\right)}{\kappa\omega_{0}}\label{eq:Sigma_yy-2}
\end{align}

From which if $\tilde{u}_{x},\tilde{u}_{y},a_{n}^{x},a_{n}^{y}$ are
substituted, Eqs. (\ref{eq:sigma_xx}), (\ref{eq:Sigma_yy}) and (\ref{eq:sigma_xy})
follow. 
As we emphasized in the main text, the sums are over positive odd
integers while for open surfaces the sums are over positive even integers.

\subsection{\label{sec:Hall-coeffcient-calculations}Hall coefficient for rectangular
lattice at limiting cases}

For the high-field limit: 
\begin{equation}
\frac{1}{R_{H}}=\lim_{B\rightarrow\infty}\frac{1}{B}\frac{\sigma_{xy}}{\sigma_{xx}\sigma_{yy}+\sigma_{xy}^{2}}
\end{equation}
Using the above equations, we obtain:

\begin{equation}
{R_{H}^{e}=-\frac{1}{\pi}\sum_{n=1}^{\infty}\frac{1}{\left(n-\frac{1}{2}\right)}sech^{2}\left[\left(n-\frac{1}{2}\right)\frac{\pi K^{'}\left(\kappa\right)}{K\left(\kappa\right)}\right]\sin\left[\left(2n-1\right)\frac{\pi u_{e}}{K\left(\kappa\right)}\right]}
\end{equation}

\begin{equation}
{R_{H}^{h}=\frac{1}{\pi}\sum_{n=1}^{\infty}\frac{1}{\left(n-\frac{1}{2}\right)}sech^{2}\left[\left(n-\frac{1}{2}\right)\frac{\pi K^{'}\left(\kappa\right)}{K\left(\kappa\right)}\right]\sin\left[\left(2n-1\right)\frac{\pi u_{h}}{K\left(\kappa\right)}\right]}
\end{equation}

\begin{equation}
{R_{H}^{o}=-\frac{1}{\pi}\frac{\sum_{n=1}^{\infty}\frac{1}{n^{2}}sech^{2}\left[\frac{n\pi K^{'}\left(1/\kappa\right)}{K\left(1/\kappa\right)}\right]\sin^{2}\left[\frac{n\pi u_{o}}{K\left(1/\kappa\right)}\right]}{\sum_{n=1}^{\infty}\frac{1}{2n}sech^{2}\left[\frac{n\pi K^{'}\left(1/\kappa\right)}{K\left(1/\kappa\right)}\right]\sin\left[\frac{2n\pi u_{o}}{K\left(1/\kappa\right)}\right]}-\frac{1}{\pi}\sum_{n=1}^{\infty}\frac{1}{n}sech^{2}\left[\frac{n\pi K^{'}\left(1/\kappa\right)}{K\left(1/\kappa\right)}\right]\sin\left[\frac{2n\pi u_{o}}{K\left(1/\kappa\right)}\right]}
\end{equation}
and we observe that in the high-field case the Hall number does not
depend on the frequency. Similarly, for the low-field case the formulae
are: 
\begin{equation}
\frac{1}{R_{H}}=\lim_{B\rightarrow0}\frac{1}{B}\frac{\sigma_{xy}}{\sigma_{xx}\sigma_{yy}+\sigma_{xy}^{2}}
\end{equation}
similarly we get:

\begin{equation}
{R_{H}^{e}=-\frac{4}{\pi}\frac{\left[\sum_{n=1}^{\infty}sech^{2}\left[\left(n-\frac{1}{2}\right)\frac{\pi K^{'}}{K}\right]\sin^{2}\left[\left(n-\frac{1}{2}\right)\frac{\pi u_{e}}{K}\right]\right]\left[\sum_{n=1}^{\infty}sech^{2}\left[\left(n-\frac{1}{2}\right)\frac{\pi K^{'}}{K}\right]\cos^{2}\left[\left(n-\frac{1}{2}\right)\frac{\pi u_{e}}{K}\right]\right]}{\sum_{n=1}^{\infty}\left(n-\frac{1}{2}\right)sech^{2}\left[\left(n-\frac{1}{2}\right)\frac{\pi K^{'}}{K}\right]\sin\left[\frac{\left(2n-1\right)\pi u_{e}}{K}\right]}}
\end{equation}

\begin{equation}
{R_{H}^{h}=\frac{4}{\pi}\frac{\left[\sum_{n=1}^{\infty}sech^{2}\left[\left(n-\frac{1}{2}\right)\frac{\pi K^{'}}{K}\right]\sin^{2}\left[\left(n-\frac{1}{2}\right)\frac{\pi u_{h}}{K}\right]\right]\left[\sum_{n=1}^{\infty}sech^{2}\left[\left(n-\frac{1}{2}\right)\frac{\pi K^{'}}{K}\right]\cos^{2}\left[\left(n-\frac{1}{2}\right)\frac{\pi u_{h}}{K}\right]\right]}{\sum_{n=1}^{\infty}\left(n-\frac{1}{2}\right)sech^{2}\left[\left(n-\frac{1}{2}\right)\frac{\pi K^{'}}{K}\right]\sin\left[\frac{\left(2n-1\right)\pi u_{h}}{K}\right]}}
\end{equation}

\begin{equation}
{R_{H}^{o}=-\frac{4}{\pi}\frac{\left[\sum_{n=1}^{\infty}sech^{2}\left[\frac{n\pi K^{'}}{K}\right]\sin^{2}\left[\frac{n\pi u_{o}}{K}\right]\right]\left[\frac{1}{2}+\sum_{n=1}^{\infty}sech^{2}\left[\frac{n\pi K^{'}}{K}\right]\cos^{2}\left[\frac{n\pi u_{o}}{K}\right]\right]}{\sum_{n=1}^{\infty}n\,sech^{2}\left[\frac{n\pi K^{'}}{K}\right]\sin\left[\frac{2n\pi u_{o}}{K}\right]}}
\end{equation}

\end{widetext}

\section{\label{sec:Side-calculation}Solution to relevant ordinary differential
equations (ODEs)}

\subsection{\label{subsec:General-setup}General setup}

Eq.(\ref{eq:Derivative-1}) is of the general form:
\begin{equation}
\frac{d\mathbf{X}}{dt}=M\left(t\right)\mathbf{X}+\mathbf{V}\left(t\right)\label{eq:Solution-1-1}
\end{equation}
First we need to solve for 
\begin{equation}
\frac{d\mathbf{X}_{0}}{dt}=M\left(t\right)\mathbf{X}_{0}\label{eq:Derivative-2}
\end{equation}
Where 
\begin{equation}
\mathbf{X}_{0}\left(t\right)=T\left(\exp\left(\int_{t_{0}}^{t}M\left(s\right)ds\right)\right)\mathbf{X}_{0}\left(t_{0}\right)\label{eq:Solve}
\end{equation}
Here $T$ stands for time ordering. Then, for Eq.(\ref{eq:Solution-1-1}), we look for solutions of the form 
\begin{align}
\mathbf{X}\left(t\right) & =T\left(\exp\left(\int_{t_{0}}^{t}M\left(s\right)ds\right)\right)\mathbf{\mathbf{Y}}\left(t\right)\nonumber \\
\frac{d\mathbf{X}\left(t\right)}{dt} & =M\left(t\right)\mathbf{X}\left(t\right)+T\left(\exp\left(\int_{t_{0}}^{t}M\left(s\right)ds\right)\right)\frac{d\mathbf{Y}\left(t\right)}{dt}\nonumber \\
 & \equiv M\left(t\right)\mathbf{X}+\mathbf{V}\left(t\right)\nonumber \\
\Rightarrow\frac{d\mathbf{Y}\left(t\right)}{dt} & =T\left(\exp\left(\int_{t_{0}}^{t}M\left(s\right)ds\right)\right)^{-1}\mathbf{V}\left(t\right)\nonumber \\
\mathbf{Y}\left(t\right) & =\mathbf{Y}_{0}+\int_{t_{0}}^{t}T\left(\exp\left(\int_{t_{0}}^{t'}M\left(s\right)ds\right)\right)^{-1}\mathbf{V}\left(t'\right)dt'\label{eq:Value}
\end{align}

\subsection{\label{subsec:Application-to-our-2}Solution of Eq. (\ref{eq:K_1})
in 3D}

We use the results of Appendix \ref{subsec:General-setup}. In
our case $\mathbf{Y}_{0}=0$ and 
\begin{equation}
\mathbf{k}_{1}\left({t}\right)=\int_{t_{0}}^{{t}}T\left(\exp\left(\int_{t'}^{{t}}M\left(s\right)ds\right)\right)\mathbf{V}\left(t'\right)dt'\label{eq:X_tau-3}
\end{equation}
where $\mathbf{V}\left(t'\right)=-e{\bf E}\left(t'\right)$
and 
\begin{equation}
M_{\alpha\beta}\left(s\right)=-e\sum_{\gamma\delta}\varepsilon_{\alpha\gamma\delta}{B_{\delta}}\frac{\partial^{2}}{\partial{k}_{\beta}\partial{k}_{\gamma}}\varepsilon\left({\bf k}_{0}\left(s\right)\right)\label{eq:Our_case-2}
\end{equation}
Then, for $t>t_{0}$:
\[
\mathbf{k}_{1}\left(t\right)=-\int_{t_0}^{t}T\left(\exp\left(\int_{t'}^{t}M\left(s\right)ds\right)\right)^{-1}\mathbf{V}\left(t'\right)dt'
\]

\subsection{\label{subsec:Application-to-our}Solution of Eq. (\ref{eq:Derivative-1})
in 3D}

Then $\mathbf{Y}_{0}=0$ and for $t>t_{0}$ 
\begin{equation}
\mathbf{k}_{1}\left({t}\right)=\int_{t_{0}}^{{t}}T\left(\exp\left(\int_{t'}^{t}M\left(s\right)ds\right)\right)\mathbf{V}\left(t'\right)dt'\label{eq:X_tau}
\end{equation}
In addition: 
\begin{equation}
\mathbf{V}\left(t'\right)=-D^{-1}\left(\mathbf{k}_{0}\left(t'\right)\right)\left[e{\bf E}\left(t'\right)+e^{2}\left({\bf B}\cdot{\bf E}\left(t'\right)\right){\bf \Omega}\left({\bf k}_{0}\left(t'\right)\right)\right]\label{eq:V_tau}
\end{equation}
and 
\begin{equation}
M_{\alpha\beta}\left(s\right)=-e\sum_{\gamma\delta}\varepsilon_{\alpha\gamma\delta}{B_{\delta}}\frac{\partial}{\partial{k}_{\beta}}\left[D^{-1}\left(\mathbf{k}_{0}\left(s\right)\right)\left[\frac{\partial}{\partial{k}_{\gamma}}\varepsilon_{M}\left({\bf k}_{0}\left(s\right)\right)\right]\right]\label{eq:Our_case-2}
\end{equation}
therefore for $t>t_{0}$ 
\begin{equation}
\mathbf{k}_{1}\left(t\right)= -\int_{t_0}^{t}T\left(\exp\left(\int_{t}^{t'}M\left(s\right)ds\right)\right)^{-1}\mathbf{V}\left(t'\right)dt'\label{eq:X_tau-2}
\end{equation}

We note that the perturbative correction in Eq. (A17), $\mathbf{k}_{1}\left(l\right)$
is linear in the external electric field as can be seen by Eqs. (D8), (D9)
and (D10). As a result, these terms are well convergent for small electric
fields. Due to the time ordered exponential in
Eq. (D9) though, there may be limitation to how large the eigenvalues of the matrix
in Eq. (D9) can be. The reason is that although there is a damping exponential $\eta\left(t;t_0\right)$
which makes various integrals convergent while the exponential in
Eq. (D10) can lead to divergences. Therefore, it is important that the damping
term is greater that the divergent one and as such the eigenvalues of the matrix
in Eq. (D9) need to be smaller then $\frac{1}{\tau_{S}\left(\mathbf{k}\right)}$.

\subsection{\label{subsec:Application-to-our-1} Solution of equations in 2D}

In a similar way we've got $\mathbf{Y}_{0}=0$ and for $t>t_{0}$
\begin{equation}
\mathbf{k}_{1}\left(t\right)=\int_{t_0}^{t}T\left(\exp\left(\int_{t'}^{t}M\left(s\right)ds\right)\right)\mathbf{V}\left(t'\right)dt'\label{eq:X_tau-1}
\end{equation}
We also have that: $\mathbf{V}\left(t'\right)=-eD^{-1}\left(\mathbf{k}_{0}\left(t'\right)\right){\bf E}\left(t'\right)$
and: 
\begin{equation}
M_{\alpha\beta}\left(s\right)=-e\sum_{\gamma}\varepsilon_{\alpha\gamma}B\frac{\partial}{\partial{k}_{\beta}}\left[D^{-1}\left(\mathbf{k}_{0}\left(s\right)\right)\left[\frac{\partial}{\partial{k}_{\gamma}}\varepsilon_{M}\left({\bf k}_{0}\left(s\right)\right)\right]\right]\label{eq:Our_case-1-1}
\end{equation}
Therefore for $t>t_{0}$ the solution reads: 
\begin{equation}
\mathbf{k}_{1}\left(t\right)=-\int_{t_0}^{t}T\left(\exp\left(\int_{t}^{t'}M\left(s\right)ds\right)\right)^{-1}\mathbf{V}\left(t'\right)dt'\label{eq:X_tau-2-1}
\end{equation}

\end{document}